# Quantifying Phase Transformations in Alloying Anodes via In-Situ Liquid Cell Hard X-ray Spectroscopy and Cryogenic Microscopy

Neil Mulcahy[a], Syeda Ramin Jannat[a], Yaqi Li[a], Tigran Simonian[a], Mariana Palos[a], James O. Douglas[a], Jessica M. Walker[b], Baptiste Gault[a,c,d], Mary P. Ryan[a], Michele Shelly Conroy[a*]

a. Department of Materials and London Centre for Nanotechnology, Imperial College London, Exhibition Road, London SW7 2AZ, U.K.

b. Diamond Light Source, Harwell Science and Innovation Campus, Didcot, UK

c. Max-Planck Institute for Sustainable Materials, Max-Planck Str. 1, 40237 Düsseldorf, Germany

d. Univ Rouen Normandie, CNRS, INSA Rouen Normandie, Groupe de Physique des Matériaux, UMR 6634, F-76000 Rouen, France

*Corresponding author: mconroy@imperial.ac.uk

## Abstract

Understanding electrochemical phenomena at complex liquid–solid interfaces requires linking real-time structural dynamics with atomic-scale interfacial chemistry. Here, we integrate operando synchrotron X-ray fluorescence/diffraction with high-resolution cryogenic electron and ion multi-model microscopy to provide a mechanistic understanding of Pt-based alloying anodes across length scales. We directly observe the initial lithiation-driven formation of $Li_2Pt$ and its evolution to a stable LiPt intermetallic phase during extended cycling via a solid-solution type reaction mechanism. Simultaneously, the solid-electrolyte interphase transitions from an unstable carbonate-rich to a stable LiF-dominated composition, confirmed by cryogenic scanning transmission electron microscopy/electron energy loss spectroscopy. Crucially, cryogenic atom probe tomography reveals spatially distinct compositional regimes within the alloy anode: a lithium-flux-limited, heterogeneous interfacial zone and a diffusion-controlled, homogeneous LiPt alloy bulk. This nanoscale compositional gradient rationalises the emergent solid-solution reaction mechanism and highlights how kinetic limitations and interface dynamics govern alloy formation and electrochemical stability. Our findings demonstrate a broadly applicable correlative framework bridging operando structural dynamics with near-atomic-resolution interfacial chemistry, advancing the rational design of durable alloy electrodes for next-generation energy storage.

# Introduction

The performance and stability of electrochemical energy storage devices are governed by dynamic processes at liquid–solid interfaces[1-3]. Here, the transport of ions and electrons drives phase transformations, the evolution of interphases, and ultimately degradation[4-6]. Capturing these phenomena with high spatial and chemical resolution remains a significant challenge, as many are highly transient, atomic-scale processes[7-9] that are sensitive to environmental exposure and susceptible to damage from irradiation, including electron beams[10-12].

Operando liquid-based platforms, such as liquid cell transmission electron microscopy (LCTEM), have enabled direct observation of certain battery processes at high temporal and spatial resolutions[8, 13, 14] while maintaining systems in their native environment. However, the presence of liquid in the cell increases incident beam scattering, fundamentally limiting the ability to acquire detailed spectroscopic information or to identify interphase compositions and electrode crystalline phases[15-17]. In the case of LCTEM, electron beam irradiation can induce various damage mechanisms such as radiolysis, further restricting critical analysis[12, 16].

To overcome these challenges, synchrotron-based hard X-ray nanoprobe techniques, including X-ray fluorescence (XRF) mapping and X-ray diffraction (XRD), offer powerful in-situ and operando tools that can provide quantitative chemical and structural information in liquid environments[18-21]. In contrast to electron microscopy, these techniques are capable of probing large liquid volumes and electrode materials with minimal scattering and substantially reduced beam damage[19, 22]. However, light elements (e.g. H, Li, C, O, F), which are central to the formation of many battery phenomena such as the solid-electrolyte interphase (SEI), are difficult to detect with X-ray methods. Further, the relatively long scan times needed to achieve sufficient signal under liquid conditions can render it difficult to follow highly transient or spatially inhomogeneous reactions.

To overcome the limitations inherent to both X-ray and electron-based characterisation techniques, correlative approaches that integrate operando or in-situ measurements with high resolution microscopy techniques are required. Methods such as cryogenic atom probe tomography (cryo-APT) and cryogenic scanning transmission electron microscopy (STEM) have emerged as powerful tools for probing electrochemical processes[9, 23]. Previous studies have successfully correlated LCTEM observations with cryo-APT analysis, capturing transient electrochemical phenomena while preserving the final interphase layer and electrode composition at high resolutions[9, 23]. These advances highlight the potential of cryogenic correlative techniques to bridge the gap between dynamic operando measurements and atomic-scale structural characterisation.

Cryo-APT provides unparalleled near-atomic-scale compositional insight into preserved interfacial regions, particularly for light elements and SEI chemistry[24, 25]. Cryo-STEM, when combined with spectroscopy techniques such as electron energy loss spectroscopy (EELS) and energy dispersive X-ray spectroscopy (EDX), offers high spatial and chemical resolution imaging of electrode morphology and local composition, enabling detailed mapping of interfacial regions and identification of nanoscale phases or SEI components[10, 26, 27]. By correlating these complementary measurements, it becomes possible to directly link structural transformations, elemental migration, and interfacial stability across multiple length scales, thereby providing mechanistic insights into alloying reactions, SEI evolution, and electrode reversibility in next-generation battery materials.

Here, we apply this combined methodology to a model platinum (Pt) alloy anode. Pt offers a simplified system in which alloying and dealloying steps are well defined and readily tracked. By contrast, technologically relevant materials such as Si, Sn, and Al exhibit far greater complexity, with larger volume changes, heterogeneous SEI formation, and complex alloying pathways. By bridging operando structural insights with near-atomistic compositional mapping, our approach establishes a transferable correlative framework that can be extended to these more challenging chemistries, providing mechanistic foundations for the design of durable next-generation energy storage systems. In this context, the Pt model system functions as a benchmark, enabling method validation and providing foundational insights for the rational design of durable, high-performance energy storage systems.

# Materials and Methods

## Synchrotron In-Situ Liquid Cell Setup

The in-situ liquid set up at the hard X-ray nanoprobe I14 beamline at Diamond Light Source (Didcot, UK) has been described in numerous reports[18, 28-30]. For this work the nanocell was composed of liquid cell electrochemical biasing nanochips which are commercially available from DENSsolutions B.V. (Delft, The Netherlands). Importantly, the commercial liquid cell electrochemical chips used here are the same as those widely employed in liquid cell TEM studies, enabling a direct comparison of the electrochemical and fluidic environments between the X-ray beamline and the TEM. This compatibility, combined with our previously demonstrated transferable liquid cell electrochemical holder design, facilitates truly seamless correlative in situ measurements between the I14 hard X-ray nanoprobe and the TEM and ensures that the operating conditions remain genuinely comparable. The bottom chip contains a three-electrode set up involving a working, reference and counter electrode made of Pt. The working electrodes are deposited on a 50 nm $SiN_x$ electron transparent membrane window with dimensions of approximately 20 µm x 200 µm[31]. Combined with an identical $SiN_x$ membrane window on the top nanochip, the electrodes of interest could be viewed.

While liquid flow is possible with this set up, it was avoided to mitigate any potential air contamination due to the air and moisture sensitive nature of the Li based electrolyte used in this experiment[2, 32, 33], $LiPF_6$ in ethylene carbonate/dimethyl carbonate (EC/DMC), supplied by Merck Life Science UK Ltd (Dorset, United Kingdom). Instead, the nanocell with electrolyte was entirely constructed within an Ar glovebox, LABmaster SP glovebox supplied by MBRAUN (Garching, Germany), and this process is illustrated in Figure 1.

The sample setup is composed of the top and bottom Si wafer nanochips, an ethylene propylene diene monomer (EPDM) o-ring, the sample holder and the casing lid. These elements are represented schematically in Figure 1 (a). The bottom nanochip containing the three Pt electrode set up was placed into the sample holder. A pipette was then used to drop cast ≈ 1 µl of electrolyte onto the bottom nanochip, as in Figure 1 (b). A photograph of this process can be seen in Figure 1 (d). The top nanochip with an O-ring was placed on top of the bottom nanochip before the two were sealed together by securely screwing the casing lid into the sample holder, as in Figure 1 (c). During this construction process it was ensured that the inlet and outlet holes on the nanochip were completely sealed to prevent any air exposure during transport from the glovebox to the beamline, and to avoid any electrolyte leakage. Due to challenges in assembling the nanocells within the glovebox, a window alignment check was performed outside.

Once the sample had been constructed it was removed from the glovebox and contact pads on the bottom nanochip were attached to external electrochemical contacts. This was subsequently connected to an external potentiostat, SP-200 potentiostat, controlled via EC-Lab®, both supplied by BioLogic (Seyssinet-Pariset, France). Details on the biasing conditions for specific experiments will be given in their respective sections. A scanning electron microscopy (SEM) image showing an example of the shape of the working electrode is shown in Figure 1 (e). Once assembled, the sample is mounted on the beamline scanning stages using a holder on a kinematic mount, as described by Kerkhof et al.[18]. A photograph of the sample mount on the beamline scanning stage can be seen in Figure 1 (f).

## X-ray Diffraction and X-ray Fluorescence

Samples were analysed using a focused X-ray beam with a spot size of 50 x 50 nm. The beamline setup is such that simultaneous XRD and XRF data could be acquired. XRF measurements were acquired via continuous raster scanning using a four-element silicon drift detector supplied by RaySpec (High Wycombe, UK) positioned in a backscatter geometry. Simultaneously, XRD measurements were recorded in transmission geometry using the Excalibur detector supplied by Diamond Light Source (Didcot, UK). All XRD measurements were taken at 18 keV, covering a Q range of 0 to 4.8 Å$^{-1}$. Specific acquisition parameters such as exposure time, scan step size and scan area will be described within the relevant results sections. More information on the beamline setup is described by Quinn et al.[29].

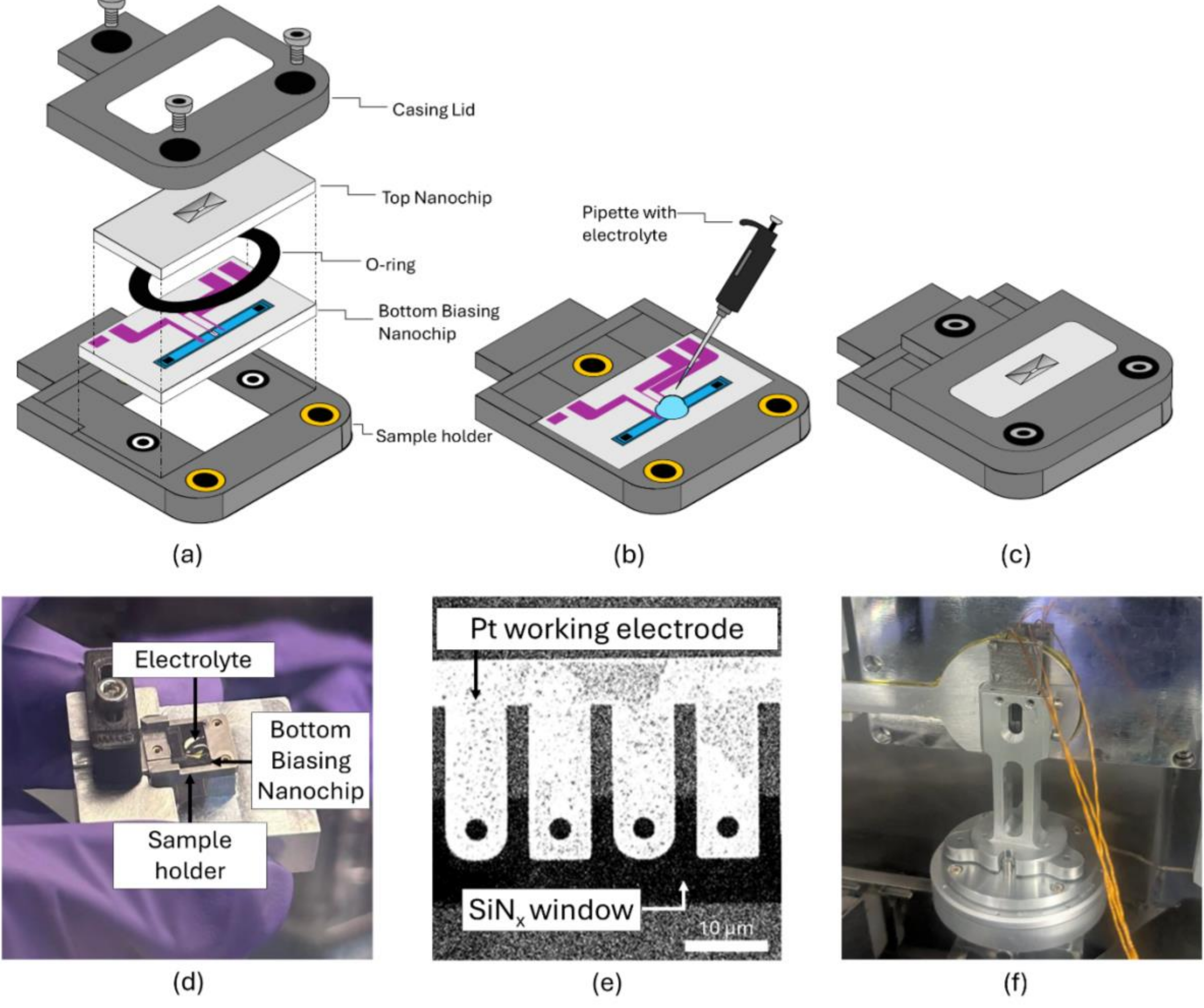

**Fig. 1:** Overview of the in situ nanocell preparation process. (a) A schematic of the contents of a fully assembled sample, comprising of the sample holder, bottom biasing nanochip, EPDM O-ring, top nanochip, and casing lid. (b) illustrates the process of drop casting Li electrolyte onto the bottom biasing nanochip using a pipette within the Ar glovebox, while (c) shows the completed closed sample with electrolyte. (d) provides a photograph of an example of the electrolyte dropcast on the nanochip. (e) shows an SEM image capturing the shape of the Pt working electrode. (f) shows the complete setup with the sample holder connected to a potentiostat on the beamline scanning stage.

To perform initial screening and quality assessment of the collected datasets, raw XRF and XRD datasets were first pre-screened using DAWN (Data Analysis WorkbeNch), supplied by Diamond Light Source (Didcot, UK). Following pre-screening, XRF datasets were extracted from processed synchrotron files in the NeXus (.nxs) format using custom Python code. Element-specific intensity maps (Pt-Lα) were normalised to the global maximum intensity and plotted as spatial distributions using the recorded X and Y scan coordinates. Resulting maps were visualised with a consistent colour scale.

Raw XRD datasets were pre-processed using an asymmetric least squares (ALS) smoothing algorithm implemented in Python to subtract the background and isolate the true diffraction signal. Data below $Q = 1$ Å$^{-1}$ were excluded to minimise low-angle noise, and the resulting baseline-corrected datasets were saved in CSV format. These were subsequently imported into Origin plotting software for manual peak fitting, where peak positions, intensities, full width at half maximum (FWHM), and coefficients of determination ($R^2$) were recorded. To identify possible phases, the fitted peak positions were compared against a reference library of simulated diffraction patterns using a custom Python script. Candidate phases were assigned based on a $\pm 0.01$ Å$^{-1}$ tolerance in q-space. All custom Python scripts used for data processing and analysis are provided as Jupyter notebooks in the Supplementary Information. Brief descriptions of each notebook's functionality are also included.

## Transfer and Freezing process

After each in-situ X-ray experiment, the sample holder was removed from the beamline and transferred to the Ar glovebox. The cell was disassembled with minimal interference to the electrolyte, and the bottom and top nanochips were placed in a gel-packing box to preserve the liquid–solid interface, as shown in Figure 2 (a).

The gel box was sealed with Parafilm, heat-sealed in a bag as in Figure 2 (b), and then transferred from Diamond Light Source to an Ar glovebox at Imperial College London. Each bottom nanochip was placed in a test tube, capped with a lid containing a small hole, and removed from the glovebox. An example of this process is shown in Figure 2 (c). Liquid nitrogen ($LN_2$) was poured through the hole to freeze the nanochip and electrolyte, with the heavier Ar atmosphere minimising air exposure, as in Figure 2 (d). The frozen nanochips could then be transferred to the cryo-stage of the PFIB/SEM for further analysis, as described by Mulcahy et al[23].

## Vacuum Cryo Transfer Module and Inert Glovebox

A vacuum cryo transfer module (VCTM) from Ferrovac GmbH (Zürich, Switzerland) was used to transfer samples between instruments, maintaining them under vacuum/inert conditions and at constant cryogenic temperatures. This module features a compact ion pump and a non-evaporable getter cartridge, maintaining pressures down to $10^{-10}$ mbar. Cryogenic temperatures can be sustained within the module via an integrated dewar of $LN_2$. The module can accommodate standard and cryo-compatible sample pucks supplied by CAMECA Inc. (Madison, WI, USA). The system contains a 500 mm wobblestick with a PEEK-insulated puck manipulator allowing pucks to be picked up or released[34].

Cryogenic sample preparation was carried out in an inert nitrogen glovebox supplied by Sylatech Ltd. (York, Uk) and is operated with typical oxygen and humidity content below 5 ppm. $LN_2$ can be delivered directly into an internal bath via a pressurised external dewar, Apollo 50 dewar supplied by Cryotherm GMBH (Kirchen (Sieg), Germany), enabling in-glovebox plunge freezing. Samples can be transferred under cryogenic and vacuum conditions to a VCTM using a vertical transfer "elevator" which can be cooled in a similar manner. The elevator can be raised into a pumpable loadlock chamber, which is directly connected to a Ferroloader docking station (Ferrovac GMBH, Zürich, Switzerland), allowing samples to be pulled into the module via a PEEK-insulated puck manipulator, maintaining the sample under vacuum and at cryogenic temperatures.

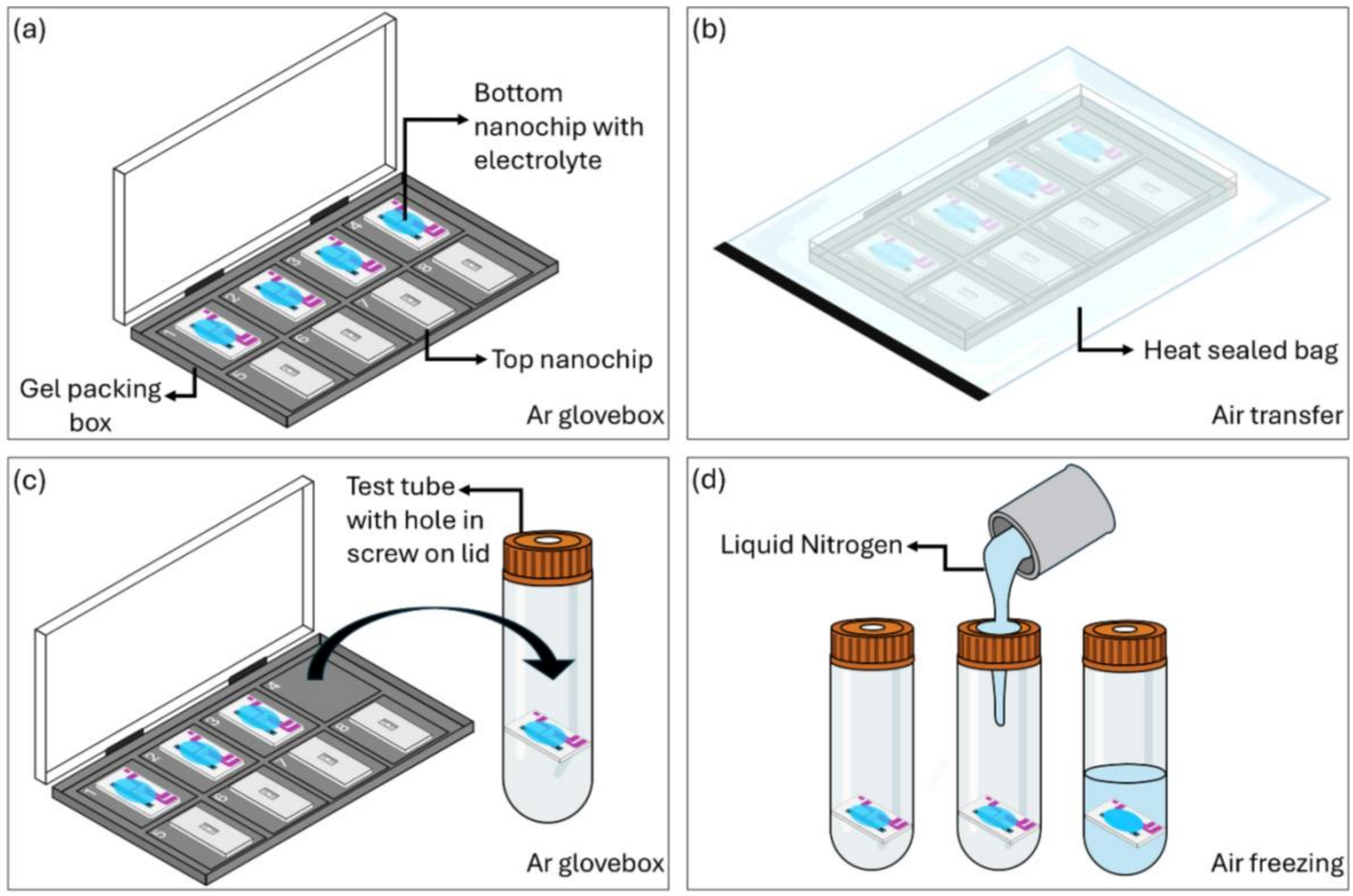

**Fig. 2:** Schematics illustrating the transfer and freezing process. (a) Bottom and top nanochips placed in a gel-packing box within an Ar glovebox at the synchrotron facility following X-ray experiments. (b) The gel box sealed with parafilm, heat-sealed in a bag, and removed from the glovebox for transfer in air. (c) Inside an Ar glovebox at the cryogenic facility at Imperial College London, the bottom nanochips containing the liquid–solid interface are placed into test tubes sealed with screw-on lids containing a small hole. (d) The test tubes are removed from the glovebox, and $LN_2$ is poured through the hole, freezing the nanochip and electrolyte.

## Plasma Focused Ion Beam/Scanning Electron Microscope with Cryogenic Capabilities

All focused ion beam (FIB)/SEM work was carried out using a Thermo Fisher Scientific Helios Hydra CX (5CX) plasma FIB system (Waltham, MA, USA), equipped with an Easylift tungsten cryo-micromanipulator and an Aquilos cryo-stage. The cryo system operates via a closed-loop flow of gaseous nitrogen cooled through a heat exchanger immersed in an external $LN_2$ dewar. This setup enabled cooling of both the stage and micromanipulator to approximately 90 K. A steady nitrogen gas flow rate of 180 mg/s was maintained to reach and sustain this base temperature. The temperature of the system, including both the stage and micromanipulator, could be controlled using a cryogenic temperature controller, Model 335 supplied by LakeShore Cryotronics Inc. (Westerville, OH, USA), in conjunction with integrated stage heaters. The instrument also includes a Ferroloader docking station, Ferrovac GmbH (Zürich, Switzerland), which allows direct transfer of cryogenic specimens from a VCTM onto the cryo stage without heating up or direct exposure to atmosphere.

For these experiments, a “dual-puck” stage baseplate supplied by Oxford Atomic (Oxford, UK) was employed, accommodating two standard cryo pucks supplied by CAMECA Instruments Inc. (Madison, WI, USA) simultaneously. Typically, one puck would contain the frozen liquid cell nanochip, while the other would contain a pre-prepared Si microarray coupon for cryo-APT specimen preparation or a molybdenum TEM lift-out grid for cryogenic TEM lamella preparation. For cryo-APT, the Si microarray coupon was prepared first at room temperature, including pre-cutting the posts at 0°, while also coating the posts in SEMGlu™, supplied by Kleindiek Nanotechnik GmbH (Reutlingen, Germany)[35]. This room temperature preparation procedure is detailed by Mulcahy et al.[36]. The FIB column was aligned at 52° relative to the SEM column.

## Cryogenic Scanning Transmission Electron Microscopy

The EleCryo double tilt vacuum transfer holder (Mel-Build, Fukuoka, Japan) was used for all STEM measurements. This holder allowed prepared TEM samples to be transferred under vacuum or inert conditions and subsequently cooled to cryogenic temperatures within the TEM instrument. Cryogenically prepared lamellae were loaded into the holder inside the nitrogen glovebox, preventing any ambient exposure throughout the entire process.

A Thermo-Fischer Scientific (Waltham, Massachusetts, United states) Spectra 300 (S)TEM at 300 kV accelerating voltage was used for all STEM measurements. This instrument is probe corrected and fitted with an ultra-high-resolution X-FEG Ulti-monochromator, a Gatan Continuum energy filter and K3 direct electron detector. Prior to any measurements, the holder was allowed to reach stable $LN_2$ temperatures (≈ 77 K) to minimise noise during acquisition. The measured screen current during STEM imaging was 17 pA with a probe converge angle of 30 mrad. EDX maps were acquired using a Bruker Dual-X detector with a dwell time of 10 µs and a measured screen current of 13 pA. Multiple frames were acquired and summed together to provide sufficient signal and improve signal-to-noise. Elemental maps were quantified in terms of weight percent (wt %) and processed using Velox supplied by Thermo-Fischer Scientific (Waltham, Massachusetts, United states).

4D-STEM datasets were acquired in nanobeam diffraction mode using a probe convergence semi-angle of 2.5 mrad over a 64 × 48 probe position grid with a step size of 4.2 nm. Diffraction patterns were recorded on a Gatan K3 direct electron detector with a camera length of 145 mm and were summed using Gatan Digital Micrograph 3.62 (Gatan Inc., Pleasanton, CA, USA) prior to further analysis.

STEM EELS datasets were acquired in energy-filtered TEM (EFTEM) mode with a camera length of 145 mm. The convergence and collection semi-angles were 40.5 and 156.5 mrad, respectively. The energy dispersion was 0.03 eV/channel with a dwell time of 0.05 s. The energy resolution, determined from the FWHM of the zero-loss peak, was 1 eV. For each area, 40 sequential frames were recorded, drift-corrected, and summed to improve the signal-to-noise ratio for subsequent data analysis in Gatan Digital Micrograph 3.62 (Gatan Inc., Pleasanton, California).

### Cryogenic Atom Probe Tomography

APT experiments were conducted using a Local Electrode Atom Probe 5000 XR (LEAP 5000 XR) instrument, manufactured by CAMECA Instruments Inc. (Madison, WI, USA). The system is equipped with a reflectron time-of-flight mass spectrometer and a Ferroloader docking station, enabling a VCTM to be directly docked to the system. Specimens introduced via a VCTM could be inserted directly into the APT analysis chamber via a “piggyback” puck, maintaining the specimen under vacuum and at constant cryogenic temperatures. The sample was analysed using laser pulsing analysis (35-45 pJ, 80-240 kHz, 1 ion per 100 pulses on average, 25k base temperature). 3D reconstructions and atom probe data analysis were completed using AP suite 6.3, a commercially available software from CAMECA Instruments Inc. (Madison, WI, USA.).

## Results and Discussion

### In-situ Synchrotron X-ray Analysis of Alloying and Interphase Evolution

#### Baseline Characterisation of the Liquid-Cell System

For each experiment, a large-area low resolution XRF scan (0.1 s exposure time, 1250 nm step size, 52 x 155 point scan area) was performed across the entire electrode surface, with a representative example shown in Figure 3 (a)(i). The region exhibiting the highest Pt-Lα signal across the electrodes corresponds to the area where the top and bottom $SiN_x$ windows on both chips overlap. For each experiment this area was chosen for analysis to maximise signal intensity. Figure 3 (a) presents a representative XRF scan (0.1 s exposure time, 50 nm step size, 37 x 94 point scan area) of this overlapping region: (ii) before and (iii) after the dropcasting of the Li electrolyte into the nanocell. A clear reduction in XRF signal intensity following the introduction of the electrolyte can be noted, underscoring the challenge of maintaining a sufficiently strong X-ray signal for tracking any changes during subsequent in-situ experiments.

XRD patterns (0.1 s exposure time, 50 nm step size, 37 x 94 point scan area) were taken of the same overlapping region for both the dry and wet electrode and is shown in Figure 3 (b). Multiple Pt reflections were observed including Pt (111), Pt (200) and Pt (220), highlighting the face-centred cubic (fcc) polycrystalline nature of the electrode (as evident in Supplementary Figure S1 (a)). The positions of these reflections remained unchanged upon the addition of the electrolyte, confirming structural stability of the Pt electrode. However, the overall signal intensity drops, likely due to increased X-ray absorption and scattering from the increased sample volume of the electrolyte. This again highlights the signal limitations during in-situ conditions. Additionally, in the regions around Q = 1.0–1.5 $Å^{-1}$ and Q = 3.75–4 $Å^{-1}$, the wet electrode showed a slight increase in background, which can be attributed to diffuse scattering from the amorphous liquid electrolyte[37]. Collectively, these findings confirm that the underlying crystallographic structure of the Pt electrode remains unchanged following electrolyte addition.

A model cyclic voltammetry (C.V.) curve for the first lithiation-delithiation cycle, obtained under a voltage range of 2 to –3 V vs. Pt at a scan rate of 0.01 mV/s, is shown in Figure 3 (c). The shape of the electrochemical curve highlights the reversibility of this process under these biasing conditions. Kim et. al[38] have previously reported a direct phase transformation on lithiation, involving the formation of $Li_2Pt$ from Pt, and a three-step delithiation process involving solid-solution behaviour for Pt alloying anodes. During the anodic sweep, two distinct peaks are observed, labelled 1 and 2 in Figure 3 (c), corresponding the stepwise phase transformation of $Li_2Pt$ to LiPt. Upon further delithiation, the system proceeds via a solid-solution reaction mechanism involving continuously varying lithium content, ultimately leading to the formation of $LiPt_2$ as a distinct phase prior to the eventual regeneration of the original Pt metal phase. This model electrochemical curve confirms that the system can be effectively biased to monitor phase transitions using this setup. Variations in biasing conditions required to optimise X-ray signal intensity will be discussed in later sections.

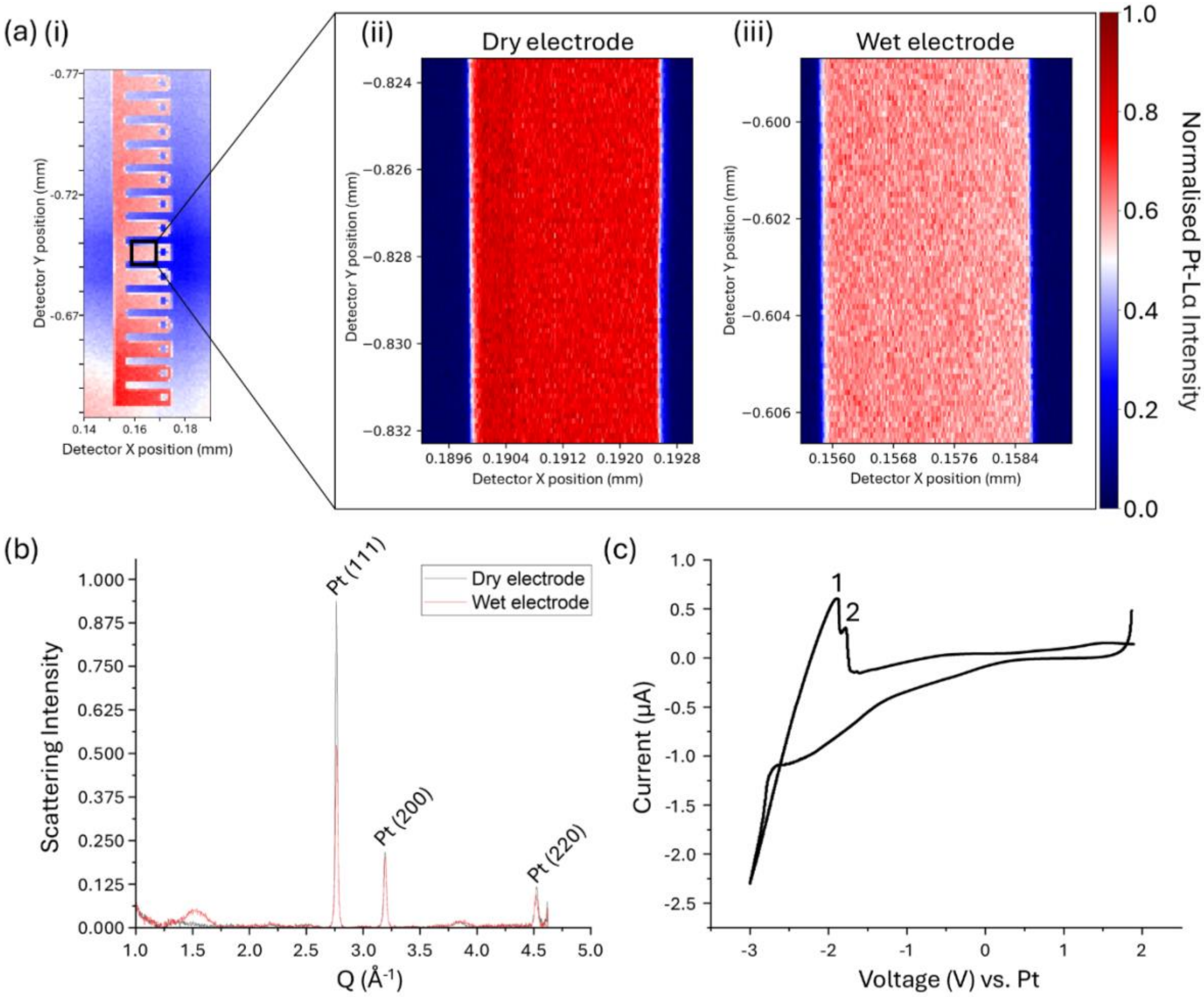

**Fig. 3:** (a) Representative XRF maps (Pt-Lα) of the electrode region showing: (i) the entire electrode area, with the portion of the electrode with maximum Pt-Lα signal highlighted by the black square, (ii) the highlighted region dry and (iii) the highlighted region with electrolyte. (b) corresponding XRD patterns collected from the same region before and after electrolyte addition, with highlighted Pt (111), (200) and (220) reflections. (c) Model cyclic voltammogram for the first lithiation–delithiation cycle of the electrode, biased between 2 and –3 V at a scan rate of 0.01 mV/s. Two peaks are labelled 1 and 2, highlighting the characteristic stepwise phase transformation during early cycling.

Together, the XRF and XRD results for both dry and wet electrodes, along with the first-cycle C.V. curve, provide essential baselines for interpreting the data presented in subsequent sections. As discussed, the addition of electrolyte to the system reduces both XRF and XRD signal intensity, highlighting the general challenge of maintaining sufficient signal under in-situ conditions. A complementary comparison in Supplementary Figure S1 using a similar nanocell setup within a TEM further illustrates characterisation challenges in liquid environments. In Figure S1 (a), a STEM image of a dry electrode is shown, where the polycrystalline grain structure of the system can be clearly imaged. In Figure S1 (b) the corresponding diffraction pattern for this electrode exhibits distinct rings, confirming the polycrystalline structure. Following the addition of electrolyte, as in Figure S1 (c), the image resolution decreases significantly, with individual grains no longer easily visible. Further, the diffraction pattern for the wet sample in Figure S1 (d) shows no clear features, emphasising the challenges of achieving high-resolution structural characterisation in liquid environments using electron microscopy techniques and highlighting the role of X-ray methods.

As a further cautionary note, Supplementary Figure S2 shows a sample likely contaminated during cell assembly or exposed to air during transport to the beamline. Sharp peaks between Q = 3.25–4 $Å^{-1}$ are evident, indicative of unintended crystalline or semi-crystalline phase formation. These peaks shifted upon biasing, emphasising the critical importance of clean, airtight sample preparation to prevent artefacts in subsequent XRD and XRF measurements.

## Lithiation: Formation of Alloys and SEI Initiation

Figure 4 (a) presents XRD datasets (0.1 s exposure time, 50 nm step size, 24 x 54 point scan area) of the Pt anode following lithiation over two separate cycles. In both cases, the formation of a Pt–Li alloy, $Li_2Pt$, is observed. The formation of $Li_2Pt$ upon lithiation has been reported previously[38]. During lithiation, the fcc Pt structure transforms into hexagonal $Li_2Pt$. Multiple reflections are detected, consistent with the polycrystalline nature of the electrode. The system was biased from 1 to –6 V at a scan rate of 0.01 mV/s, as shown in Figure 4 (b). For the XRD scan, the potential was held at –6 V for 130 seconds to generate a sufficiently strong signal for reliable data acquisition. The current remained stable throughout this hold, as shown in Figure 4 (c). The potential implications of these aggressive cycling conditions are discussed further below.

During the initial lithiation cycle, in addition to the $Li_2Pt$ alloying peaks, a number of additional reflections are detected, notably Li $(1\bar{1}\bar{1})$, Pt (111), and $Li_2CO_3$ (020). The presence of metallic Li likely indicates plating on the anode surface, which may include dendritic growth. This behaviour is common during early lithiation cycles, especially under aggressive intermediate overpotentials or when the SEI is incomplete or unstable. The formation of Li "cauliflower" deposits under these conditions, consistent with uneven Li nucleation and incomplete SEI formation, has been reported previously[9, 13]. However, determining the exact source of the Li signal is challenging based on XRD alone. The persistence of the Pt (111) reflection during the first cycle suggests that the lithiation capacity was insufficient to fully alloy the Pt anode at this stage. In contrast, during the second lithiation cycle, the Pt reflection disappears, while $Li_2Pt$ alloying peaks exhibit increased scattering intensities, indicating a more complete alloying process.

A $Li_2CO_3$ (020) reflection is detected in both cycles, reflecting ongoing formation of the SEI layer via electrolyte decomposition. $Li_2CO_3$ has been reported as a major component of the inner SEI layer of Pt alloying anodes during early cycling in previous studies[9, 38], consistent with the results obtained here. It has been shown previously that the catalytic properties of metals like Pt accelerate the decomposition of electrolyte solvents such as ethylene carbonate, particularly in the presence of trace water or HF, leading to more pronounced formation of $Li_2CO_3$ and LEDC as initial SEI components[39, 40]. Interestingly, the intensity of the $Li_2CO_3$ peak decreases significantly during the second lithiation cycle. This reduction may reflect the limited stability of $Li_2CO_3$ within the SEI. Previous work has shown that $Li_2CO_3$ can be reduced to $Li_2O$ and gaseous species such as CO or $CO_2$ under aggressive lithiation conditions or charging conditions[41, 42]. SEI cracking and mechanical instability during early cycles are also commonly observed in high-capacity alloying anodes such as Si, where large volume changes induce fracture and reformation of the SEI layer, particularly when the electrode is not fully passivated during early cycling[9, 43].

It is important to address the aggressive cycling conditions and the impact these could have on any observed phenomena. In contrast to the baseline electrochemical curve shown in Figure 3 (c), the system needed to be driven from –3 V to –6 V to generate significant alloying information. This shift can be attributed to several factors related to the limited electrolyte volume within the cell. First, concentration polarisation arises because the confined electrolyte volume causes rapid local depletion of $Li^+$ ions and changes in the local cosolvent ratio near the Pt working electrode, even at moderate current densities[44]. To overcome this depletion and sustain lithiation, a larger overpotential is required. Second, slow replenishment of $Li^+$ ions from the bulk electrolyte exacerbates this effect. Due to the limited electrolyte volume, diffusion pathways are constrained, reducing the ability of the bulk to replenish ions at the electrode surface, which increases concentration overpotential[45, 46]. Finally, electrolyte resistance plays a key role. Despite reasonable ionic conductivity, the small electrolyte volume can exhibit higher overall ohmic resistance, especially if the ionic path length is long or cross-sectional area is restricted, leading to increased voltage losses during cycling[47].

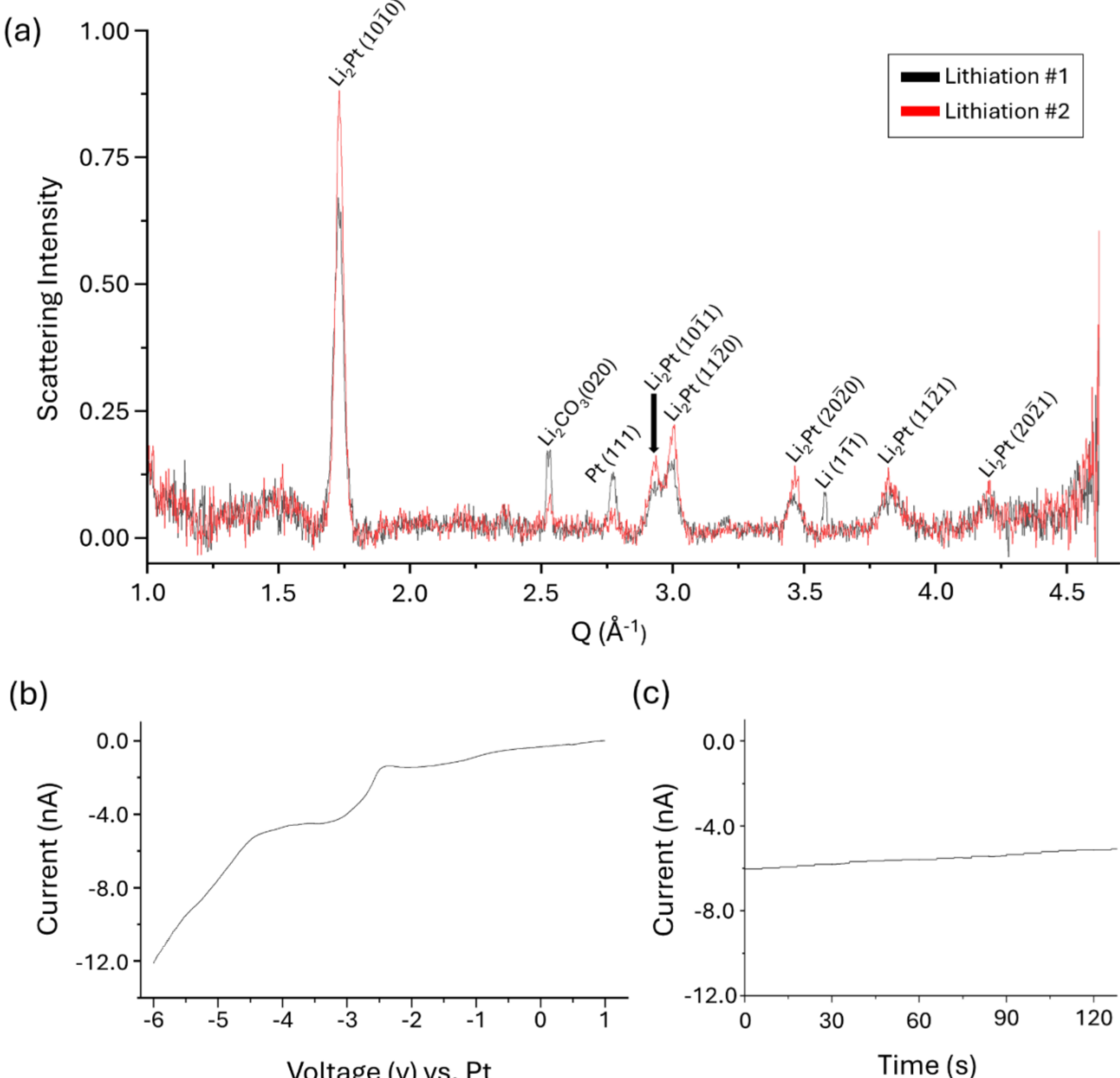

**Fig. 4:** (a) XRD patterns of the Pt anode following lithiation over two cycles with the first lithiation shown in black and the second lithiation in red, with notable reflections highlighted. (b) electrochemical biasing conditions used during XRD acquisition: (i) voltage sweep from 1 to –6 V at 0.01 mV/s and (ii) the current response vs. time during the 130 s hold at –6 V.

In addition to these volume-related effects, gas evolution during cycling may generate bubbles within the confined electrolyte, which could partially block electrode surfaces and restrict mass transport[48, 49]. This is particularly detrimental in such confined nanocell volumes where the effective transport cross-section is already limited. Solvent evaporation arising from imperfect sealing of the electrochemical cell may also contribute to reducing electrolyte availability and contribute to sluggish ion transport[50, 51]. As Pt-based alloying phenomena are capacity-dependent, as reported by Kim et al.[38], reaching the critical capacity for alloying necessitates increasing the applied potential to compensate for these limitations. Despite pushing to –6 V in the first lithiation cycle, Pt reflections persist, highlighting this challenge.

The effect of holding the voltage at –6 V for 130 seconds warrants careful consideration. The stable current during this hold indicates a steady state lithiation process was established. This stability suggests that any potential $Li^+$ ion depletion near the electrode was effectively mitigated, and that both electrolyte conductivity and ionic transport were sufficient to sustain the reaction. Crucially, maintaining this potential facilitated the controlled and continuous formation of Pt–Li alloys for analysis. While such stability is not inconsistent with parasitic side reactions or Li plating occurring in parallel, the absence of current fluctuations suggests that abrupt processes such as rapid Li plating were not dominant. Moreover, the relatively short hold time should minimise the extent of any competing steady-state side reactions. Taken together, the results support the conclusion that the system remained reasonably stable under these aggressive cycling conditions.

While these conditions are well beyond the baseline potential range, they were necessary to overcome the intrinsic mass-transport and resistance limitations of the confined electrolyte environment. Importantly, the combination of a controlled –6 V hold and stable lithiation current ensured that these conditions were sufficient to drive the formation and detection of the targeted Pt–Li alloying phases with sufficient X-ray signal without inducing electrode failure or irreversible damage to the system. This balance allowed for clear observation of the alloying mechanisms while preserving the overall structural and electrochemical integrity of the MEMS cell.

### Delithiation: Reversible Dealloying and Solid-Solution reaction

Following the successful observation of lithiation alloying peaks and SEI behaviour of the Pt electrode, the system was returned to 0 V and a subsequent XRD dataset (0.1 s exposure time, 50 nm step size, 24 x 54 point scan area) was acquired, as shown in Figure 5 (a). Comparing the lithiated XRD pattern (black line) with the delithiated pattern (red line) reveals a complete transformation in the alloying reflections. Notably, all $Li_2Pt$ peaks disappear upon delithiation, replaced by a broad and prominent hexagonal LiPt (0001) reflection.

Transitioning from the $Li_2Pt$ phase to LiPt (0001), we observe significant peak broadening, which is indicative of increased lattice strain and reduced crystallite domain size during delithiation. This broadening may also reflect mechanical stresses and microstructural changes arising from repeated alloying and dealloying cycles. Literature reports that residual lithium can become trapped within the Pt lattice[9, 38], hindering a full return to the original Pt structure. Incomplete delithiation may also result from kinetic barriers associated with strain and Li diffusion limitations, as evidenced by the persistence of LiPt reflections after cycling. The broadening and persistence of LiPt peaks also indicate a transition in the dealloying mechanism. This evolution is illustrated in Figure 5 (b), which compares baseline electrochemical data against that of the electrode after more than twenty cycles, where a clear merging of anodic peaks signifies the shift from discrete phase transformations to a more solid-solution-like reaction mechanism. These structural evolutions not only influence lithium diffusion kinetics but also have important consequences for the mechanical stability and electrochemical performance of Pt-based alloy electrodes during cycling.

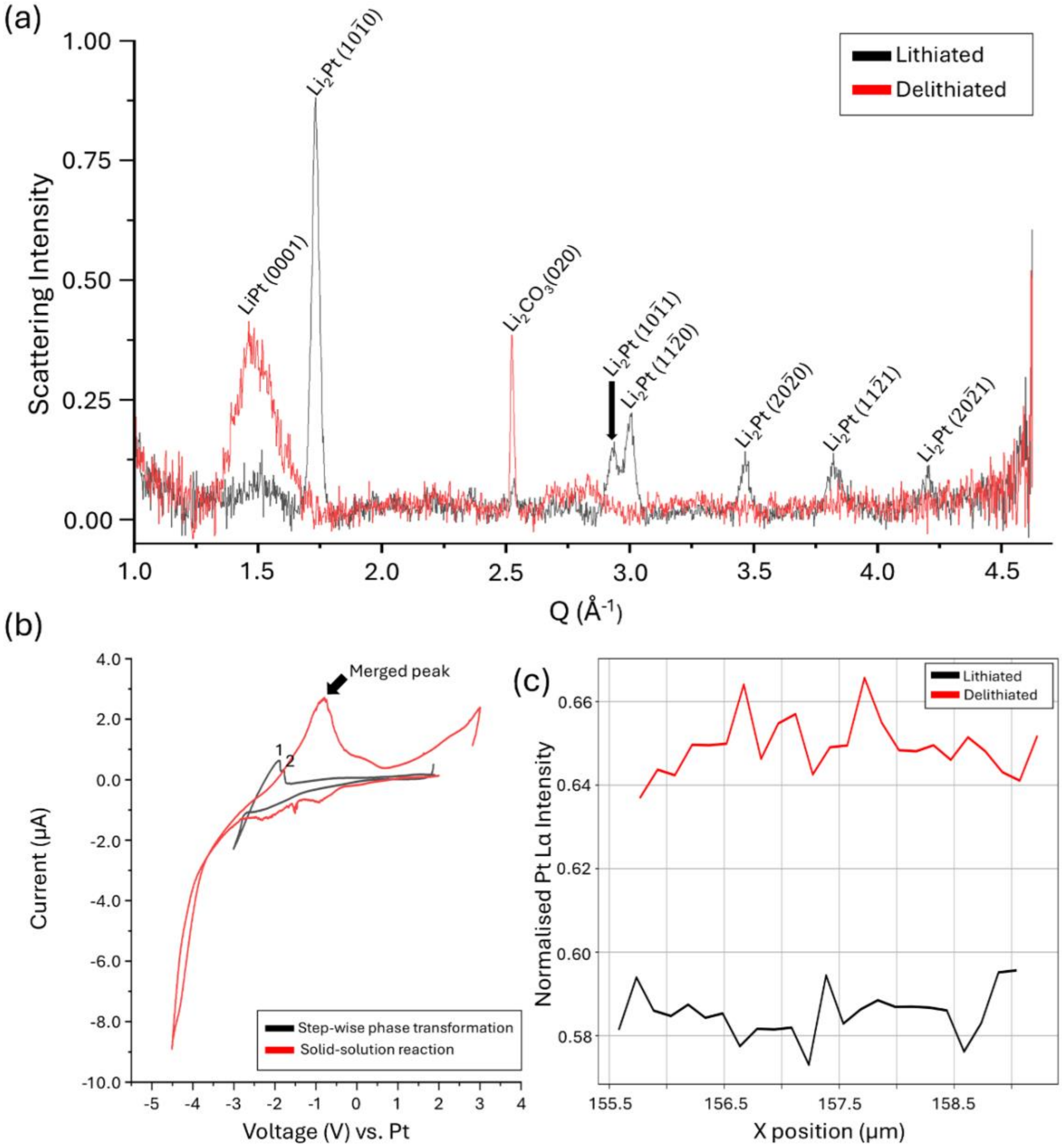

**Fig. 5:** (a) XRD patterns of the electrode in the lithiated (black) and delithiated (red) states. (b) Comparison of electrochemical profiles from the initial baseline and after more than 20 cycles. With extended cycling, the anodic peaks labeled 1 and 2 merge into a single peak. (c) Plot of normalised Pt–Lα intensity versus x-position for the lithiated (black) and delithiated (red) states.

In addition to the LiPt (001) reflection, the $Li_2CO_3$ (020) peak also persists and shows a significant increase in intensity compared with the lithiated state. These changes could arise from repeated volume changes and mechanical stresses during alloying and dealloying which could amplify the XRD signal without necessarily altering the actual thickness or composition of the SEI. Structural rearrangements within the anode during delithiation may have also exposed fresh surfaces to the electrolyte, promoting further SEI formation and thereby increasing the observed $Li_2CO_3$ signal. These observations suggest a dynamic and evolving SEI that responds to cycling, with potential implications for the long-term stability and performance of the Pt-based anode.

Simultaneous XRF measurements (0.1 s exposure time, 50 nm step size, 77 x 178 point scan area) of both the lithiated and delithiated states further highlight the ongoing microstructural evolution in terms of elemental density. While the spatial distribution of the Pt–Lα signal showed no obvious change during early cycle lithiation and delithiation (see Supplementary Figure S3), a clear difference in intensity was observed. As shown in Figure 5 (c), the delithiated state consistently produced a higher XRF signal than the lithiated state. This contrast primarily reflects changes in local elemental density resulting from electrochemical cycling. Delithiation removes Li from the Pt lattice, causing a volume contraction that increases the local density of Pt atoms within the analysis volume, thereby enhancing the detected XRF signal. Conversely, during lithiation, lattice expansion due to Li incorporation reduces the local Pt density, leading to a lower signal. Such lattice expansion during lithiation has been directly visualised in real time in previous studies[9, 13], lending further support to the structural interpretation presented here.

## Long term cycling: SEI evolution and electrode redistribution

To evaluate the durability and practical viability of this system as a battery anode, the same electrode was subjected to a total of 50 charge–discharge cycles (0.01 mV/s, 3 V to -4/4.5 V), with intermittent simultaneous XRF and XRD scans (0.1 s exposure time, 200 nm step size, 37 x 194 point scan area) performed after selected cycles. Figure 6 (a) presents the combined XRD datasets following a number of these cycles. Within the datasets, the $Li_2CO_3$ (020) reflection persists albeit diminishing with increasing cycles. Additionally, an emerging SEI reflection corresponding to LiF (200) is observed. LiF is well-established as a key stabilising component of the inner layer of the SEI in many lithium-ion battery studies[2, 52, 53] and has also been reported as part of the SEI on Pt-based alloy anodes[38].

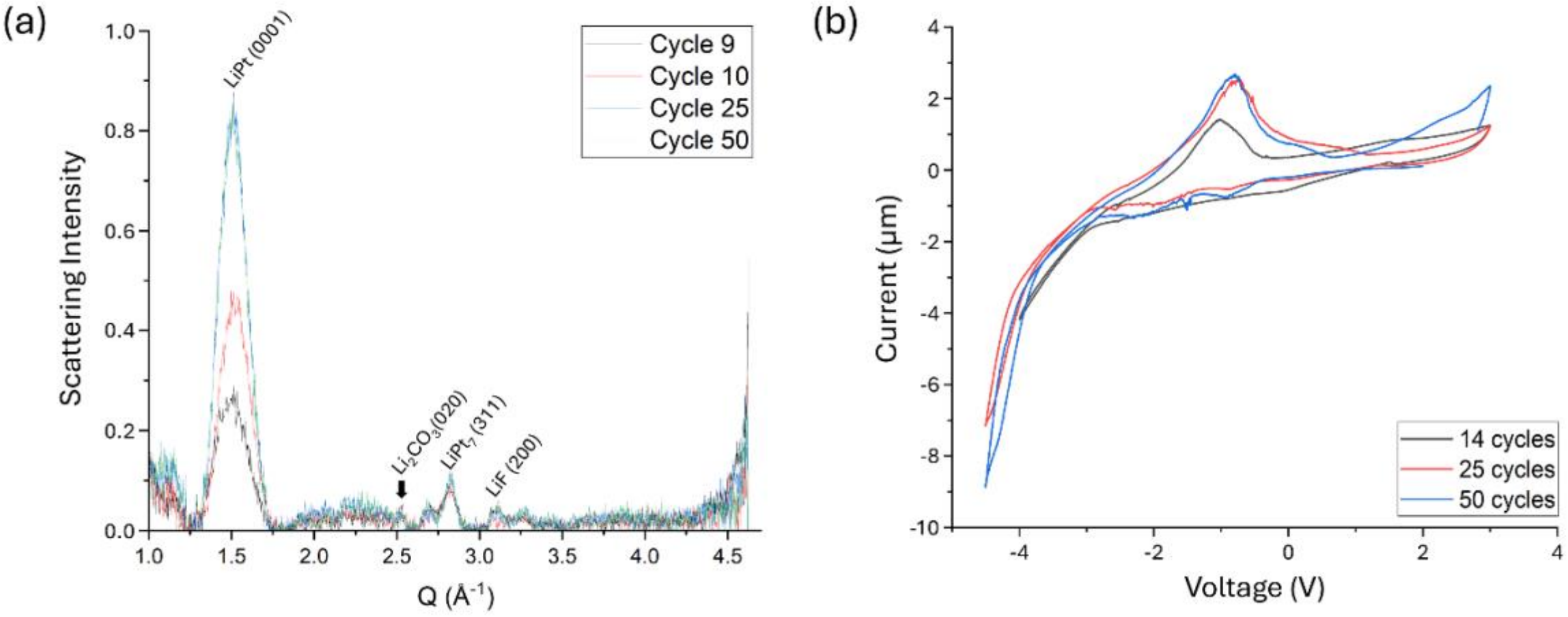

**Fig. 6:** (a) XRD patterns collected after selected cycles during 50 charge–discharge cycles, with reflections of note highlighted (b) C.V. curves at various numbered cycles.

Figure 6 (a) shows the continued persistence of the LiPt (0001) reflection even following extended cycling. Notably, the scattering intensity of this peak gradually increases before stabilising. This increase in intensity likely reflects structural changes within the electrode, such as increased crystallinity or improved phase homogeneity, factors known to enhance the diffraction signal, as observed in other in-situ X-ray studies of alloy anodes[54-56]. Concurrently, the LiPt (0001) peak shifts from approximately 1.490 to 1.510 $Å^{-1}$ in q-space denoting lattice contraction, pointing towards subtle changes in the lattice parameters, which can be attributed to strain relaxation, phase reordering, or gradual changes in Li content within the alloy, consistent with prior reports on phase evolution in alloy electrodes during Li insertion[57, 58]. Additionally, the emergence of a $LiPt_7$ (311) phase is observed with cycling, suggesting further phase evolution and increasing complexity in the alloying mechanism. Following this period, the system reaches a steady state, where both the Q-value and peak intensity of the LiPt (0001) phase remains consistent, signifying stabilisation of the electrode structure and lithium distribution. These observations highlight an initially dynamic cycling evolution, followed by sustained structural stability, which is critical for the long-term performance of the Pt-based anode.

In Figure 6 (b), the electrochemical biasing curves over multiple cycles are presented. A single peak in the anodic part of the cycle persists throughout, consistent with a solid-solution type reaction mechanism. With continued cycling, this peak gradually shifts toward lower voltages, reflecting increasing polarization/overpotential. Despite this shift, characteristic lithiation and delithiation features remain consistent after extended cycling, suggesting that electrochemical activity is maintained, albeit with evolving kinetics. Additionally, the anodic peak intensity increases over time, which may indicate improved utilisation of the electrode material or progressive changes in reaction pathways[59].

Figure 7 presents the XRF datasets of the Pt electrode over extended cycling. Over time, a pronounced change is observed in both the distribution and intensity of the Pt-Lα signal. After ten cycles, there is a notable decrease in Pt-Lα intensity across the electrode, potentially indicating increased stability of the LiPt phase. By fifteen cycles, significant redistribution of the Pt signal occurs, with distinct voids appearing within the electrode structure and Pt migrating towards the edges and corners. The redistribution of the Pt and the formation of these voids intensifies with further cycling.

As mentioned earlier, continued lithiation and delithiation induces significant volume changes and mechanical stresses within the electrode[9]. Further to this, it has been shown that Pt may migrate as a result of stress-driven diffusion or electrochemical dissolution and re-deposition processes, resulting in the observed redistribution[60, 61]. The redistribution of Pt towards edges and corners of the electrode, as in the red areas in Figure 7 cycle 16-50, may be influenced by intensified local electric fields at these geometrical features[13]. Such enhanced fields can accelerate Pt dissolution and promote its migration and redeposition preferentially at edges and corners. Combined with mechanical stresses from repeated cycling, this contributes to the evolving electrode morphology and highlights the challenges in preserving electrode stability over long-term operation.

Importantly, despite these morphological changes and observed redistribution of Pt, no significant degradation is observed in the electrochemical performance, as evidenced by the consistent C.V. curves in Figure 6 (b). This stability likely arises from the persistence of the active LiPt phase, ensuring consistent electrochemical activity throughout cycling, and the emergence of the stabilising LiF SEI component. While this redistribution and mechanical stress from constant volume expansion should impact the integrity of the fragile SEI layer, it appears to remain effective due to its dynamic nature, continuously breaking down and reforming to preserve electrode passivation.

The SEI evolution observed in this study involving early-cycle $Li_2CO_3$ formation followed by LiF, is consistent with a dynamic, self-renewing interphase. As mentioned, $Li_2CO_3$ has been shown to be unstable under repeated cycling, while LiF accumulation is frequently reported as a stabilising component that lowers interfacial impedance[52, 53, 62]. Similarly, the transition from stepwise alloying-dealloying pathway to a solid-solution type reaction mechanism pathway centred on LiPt likely reflects a mechanistic shift toward phase regimes that better accommodate strain and support reversibility[19, 38]. Together, these trends highlight that long-term stability arises from continuous renewal of the interphase, stabilising electrochemical cycling of the alloy through a solid solution type reaction mechanism. This mechanism is expected to extend to more complex alloy and conversion anodes.

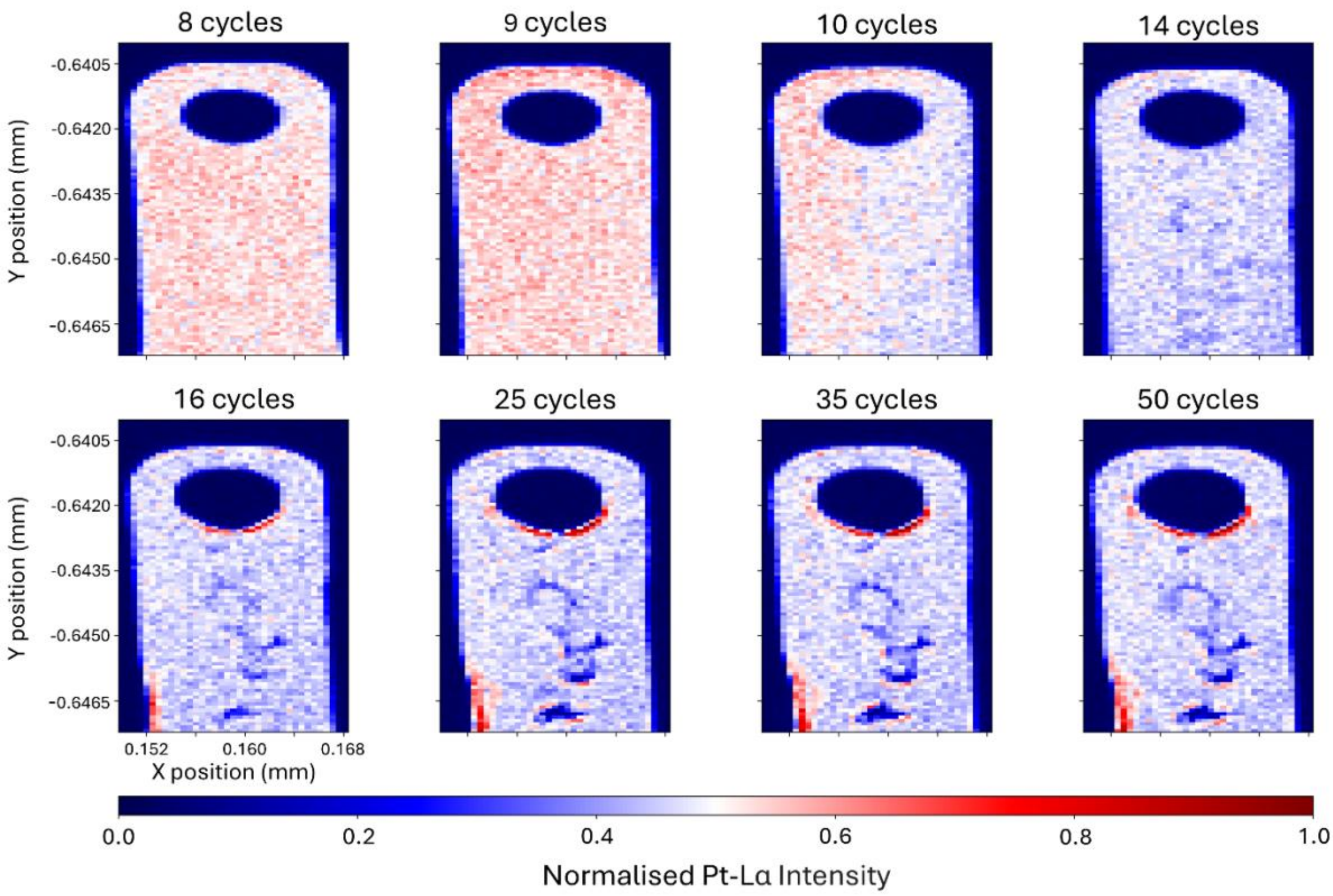

**Fig. 7:** Normalised Pt–Lα XRF intensity maps after various cycles, showing changes in Pt distribution and intensity across the electrode surface.

## Correlative Cryogenic Microscopy

### Cryogenic FIB/SEM Imaging and Cryogenic Sample Preparation of Cycled Electrodes

Following the in-situ liquid cell synchrotron XRD and XRF experiments, the MEMS nanochips were frozen and transferred to the cryo stage of a PFIB/SEM for further analysis. Figure 8 (a) shows an SEM image of a preserved section of the frozen liquid–solid interface between the electrode and the electrolyte. A thick electrolyte layer covers several electrodes which have been highlighted with white arrows. The thickness of the liquid limits detailed structural analysis. However, this image demonstrates that the liquid–solid interface was successfully preserved throughout transfer and freezing from the synchrotron to the cryogenic PFIB/SEM.

Supplementary Figure S4 provides an example of attempts made to create a lift out from this electrolyte covered region. Milling through the window revealed a thick layer of electrolyte covering the ROI, which complicated identifying the location of a suitable electrode. $SiN_x$ membrane window bulging and evidence of delamination is also evident. This may have resulted from mechanical stresses induced by repeated volume changes of the Pt electrode during aggressive lithiation and delithiation cycling. While this could potentially affect the chip's mechanical and interface integrity, the extent of the impact is unclear based on these observations alone.

As an alternative a region where the electrode was partially uncovered was used for analysis, allowing direct observation of the structural impact of long-term cycling, as shown in Figure 8 (b). Multiple voids are evident across each of the electrodes, as well as accumulation of material along the edges, most likely Pt, as highlighted in the XRF data shown in Figure 7. The electrodes look noticeably thinner, in comparison to the uncycled example in Figure 1 (e). While the electrochemical performance of the electrodes does not diminish with cycling as shown previously, the structural damage that occurs to the electrode itself is concerning, particularly for industrial relevant applications where electrode stability is key.

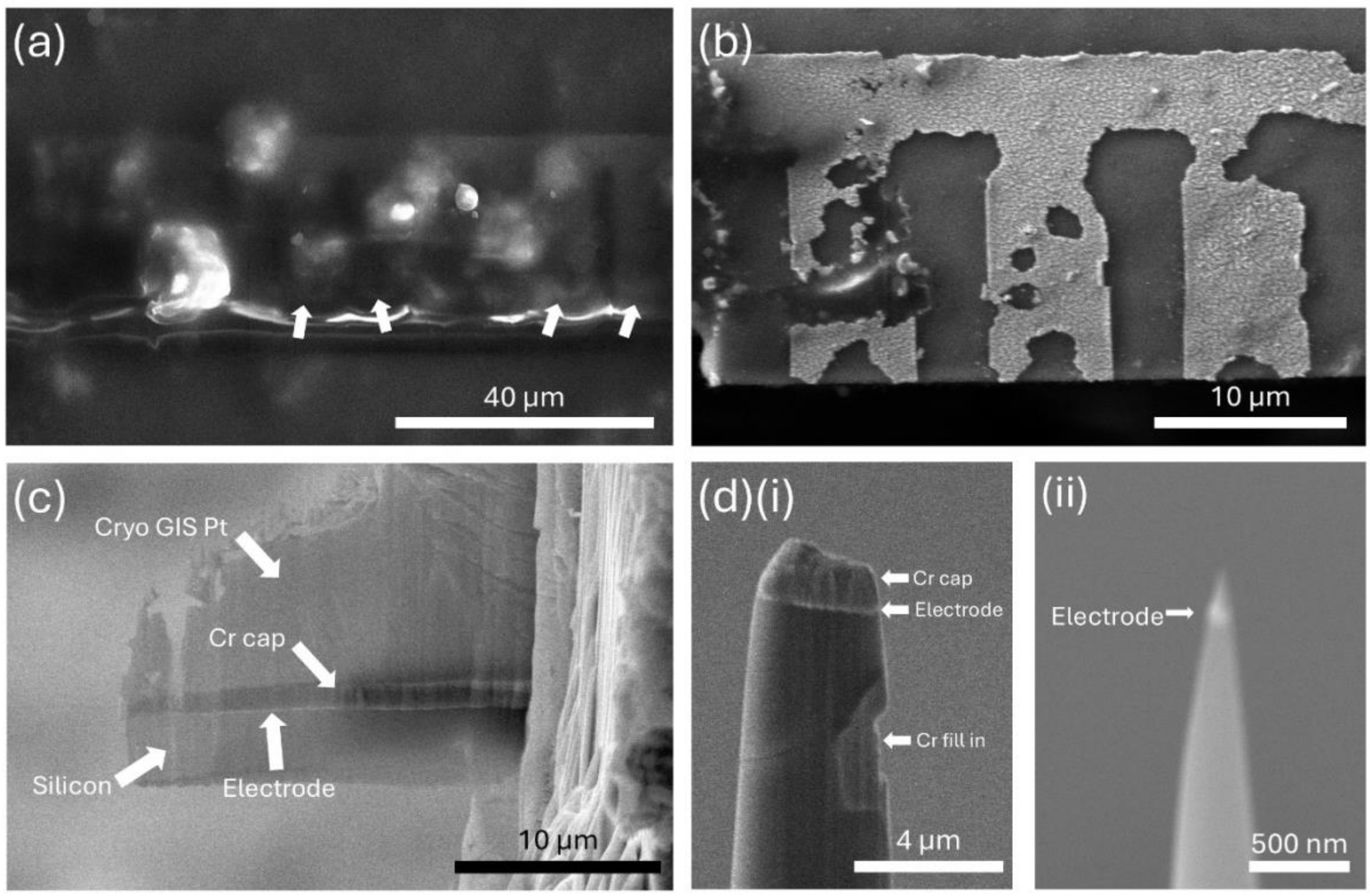

**Fig. 8:** (a) SEM image (30 kV, 1.6 nA) of electrodes, highlighted by the white arrows, buried beneath the electrolyte layer. (b) SEM image (10 kV, 0.4 nA) of a partially uncovered electrode after long-term cycling. (c) SEM image (20 kV, 1.6 nA) of a TEM lamella with visible Cr protection cap, cryogenic GIS Pt cap and electrode prepared using Ar plasma. (d) SEM images (10 kV, 0.4 nA) of the cryo-APT sample preparation: (i) lifted-out electrode on post with visible Cr protection cap and Cr fill in, and (ii) final thinned needle specimen containing the electrode.

To further understand the impact of long-term cycling on the structure of the electrode and composition of the SEI at higher resolution length scales, a portion of the exposed electrode was lifted out and prepared into both a TEM lamella and a specimen for cryo-APT analysis, as detailed for MEMS based nanochips by Mulcahy et al.[23]. Figure 8 (c) presents an example of a cryogenic TEM lamella prepared using Ar plasma with noticeable protection layers and captured electrode interface. The TEM lamella was transferred back into the glovebox via the cryogenic vacuum transfer suitcase. Then the TEM lamella was mounted into a Melbuild vacuum transfer TEM holder and the tip retracted before removing from the glovebox to avoid the sample being exposed to ambient surroundings. The holder was then placed into the TEM and the sample tip was only inserted when the rod was in the TEM column and thus below $10^{-8}$ mbar vacuum environment. Figure 8 (d)(i) presents an example of a lifted-out electrode with a deposited Cr protection layer, while (ii) presents the final thinned needle-shaped specimen using Xe plasma, with noticeable electrode captured within the tip. Specific parameters and methods for the cryo-APT sample preparation are addressed in Supplementary Figure S5.

## Cryo STEM EDX and EELS mapping of Solid-Electrolyte interphase

Figure 9 (a) presents a high-angle annular dark field (HAADF) STEM image and correlative EDX map of F, Si, Cr, Pt of the prepared TEM lamella collected at cryogenic temperatures. Also supplementary Figure S6 displays elemental maps for several detected elements including Pt, Si, Cr, N, and Xe. From these maps, it is evident that the Pt electrode on the $SiN_x$ window has been capture in cross section. Notably, a distinct region between the Cr layer deposited as FIB protection and the Pt layer shows a marked absence of detectable species, which likely corresponds to the SEI layer. This region appears non-uniform and in some areas is entirely absent, as seen in supplementary figures S6-S9, suggesting that even after extended cycling, complete passivation of the Pt electrode surface remains challenging. Furthermore, the electrode exhibits a notably uneven morphology, with signs of delamination visible in additional cryo-STEM images in Supplementary Figure S7, underscoring the effects of prolonged cycling on surface integrity and morphology. The structure of the Pt electrode is also clearly polycrystalline as seen in the Fourier transform of Figure 9 (c) high resolution STEM image and (d) 4D-STEM was acquired to confirm the polycrystalline structure of the electrode.

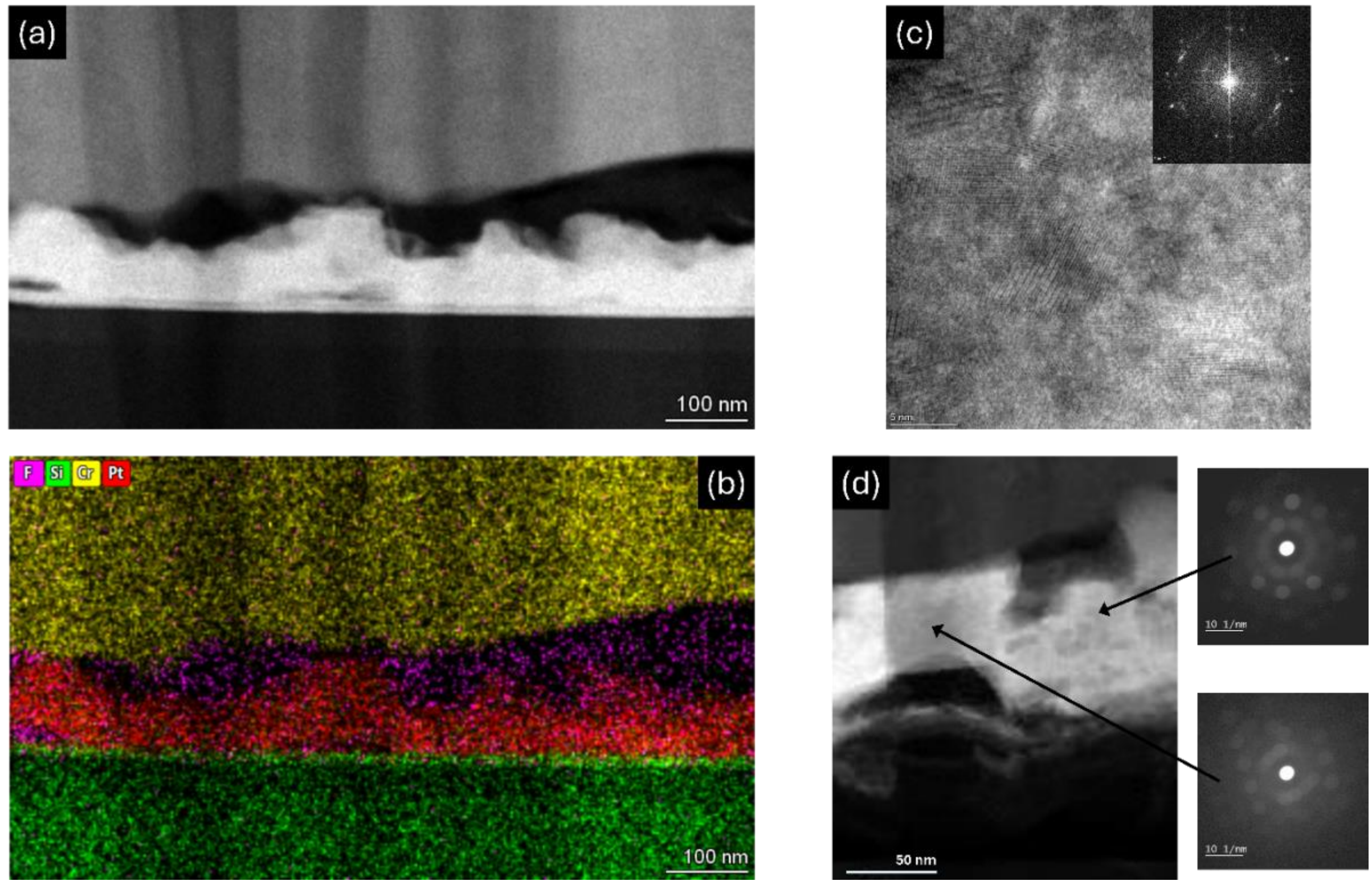

**Fig. 9:** (a) HAADF STEM image of the prepared TEM lamella(b) EDX map of the region in (a) containing F (pink), Si (green), Cr (yellow) and Pt (red) signals. EDX maps of the same region displaying (c) lattice resolution STEM with Fourier transform insert of image, (d) STEM image from 4D-STEM data set with diffraction patterns shown from different regions of the area imaged.

The STEM EDX maps of Figure 9 (a) and supplementary Figure S6 reveal that the SEI region predominantly involves F-containing species, with a notable absence of C and a reduced O signal. This suggests that the SEI is largely composed of the stabilising LiF component, while the initially abundant $Li_2CO_3$ has largely disappeared. To corroborate this observation, cryo-EELS analysis was conducted on the SEI region and is presented in Figure 10. The cryo-EELS spectrum in Figure 10 (a) exhibits a deviation from the expected Li K-edge at 55 eV, displaying a characteristic double-peak feature between ≈60–75 eV. Comparison of this peak shape with previously reported SEI components indicates a close correspondence with a spectrum characteristic of LiF[63-65]. Mapping this signal onto the ROI in Figure 10 (c) demonstrates that the LiF signal is localised exclusively within the region between the Pt and Cr layers described previously. Supplementary Figure S8 presents a high-loss range containing O edges, which again confirms the near-complete absence of oxygen-containing species within the SEI. Collectively, these observations provide evidence that the SEI evolves from early-stage, unstable carbonate species to a more robust, LiF-dominated composition, underpinning the enhanced long-term cyclability of the electrode.

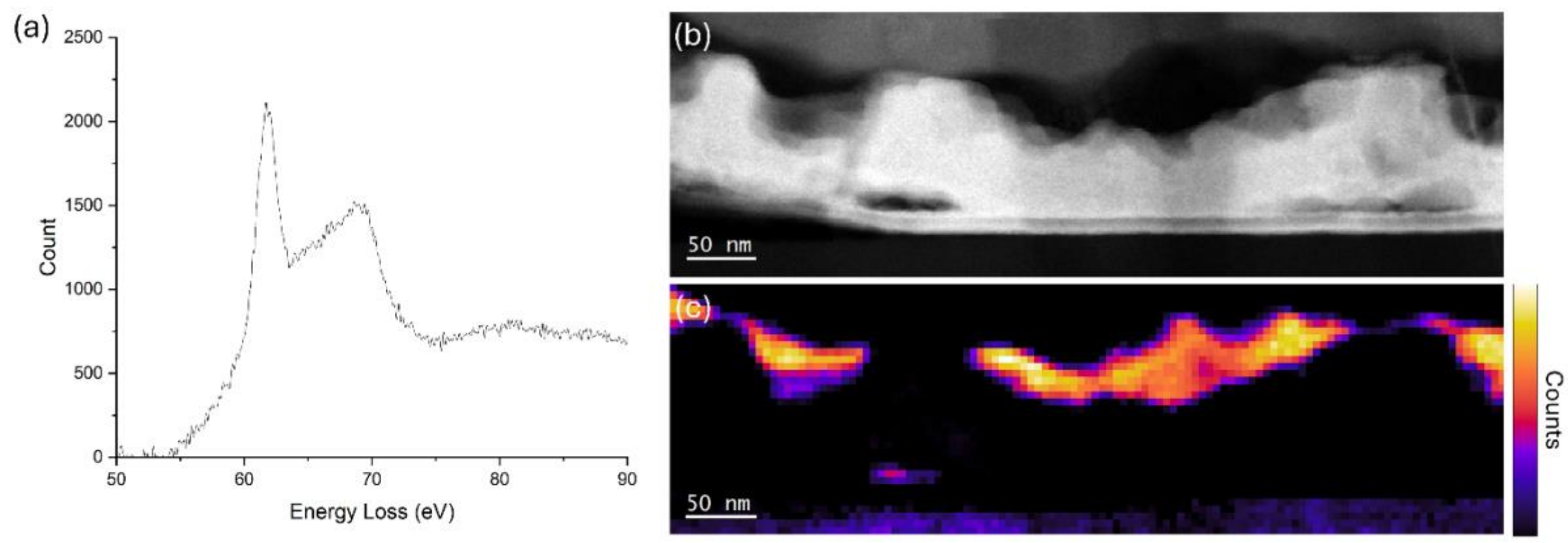

**Fig. 10:** (a) EELS spectra showing energy loss (eV) versus counts over the 50–90 eV range. (b) HAADF-STEM image of the ROI. (c) Map of the relative EELS signal intensity within the same ROI.

To further elucidate the impact of alloy formation and cycling on the anode structure, cryo-EELS analysis was conducted on a region containing the electrode, as shown in Figure 11 (a). The spectra reveal a pronounced absence of Li-related signals, with the only significant edges corresponding to Pt–O (52 eV) and Pt–N (71 eV). Mapping the EELS signal onto the ROI confirms that the detected signal is confined entirely to the electrode region previously identified by cryo-EDX in Figure 9. Although XRD analysis (Figure 6(a)) indicates that the electrode should contain a LiPt alloy, capturing Li signals within the electrode proved challenging under these conditions.

Supplementary Figure S9 provides a position-intensity map of the relative inelastic mean free path across the cryo-HAADF-STEM image in Figure 10 (c). This map reveals significant variations in sample thickness, particularly between the SEI and the electrode, which may result from differences in milling rates between the beam-sensitive SEI components and the dense Pt-alloying anode during sample preparation. These variations complicate the preparation of a uniformly thin lamella, making simultaneous detection of Li in the electrode and reliable characterisation of the SEI difficult. To overcome these limitations and more accurately determine compositional variations within the anode, cryo-APT was performed on the electrode.

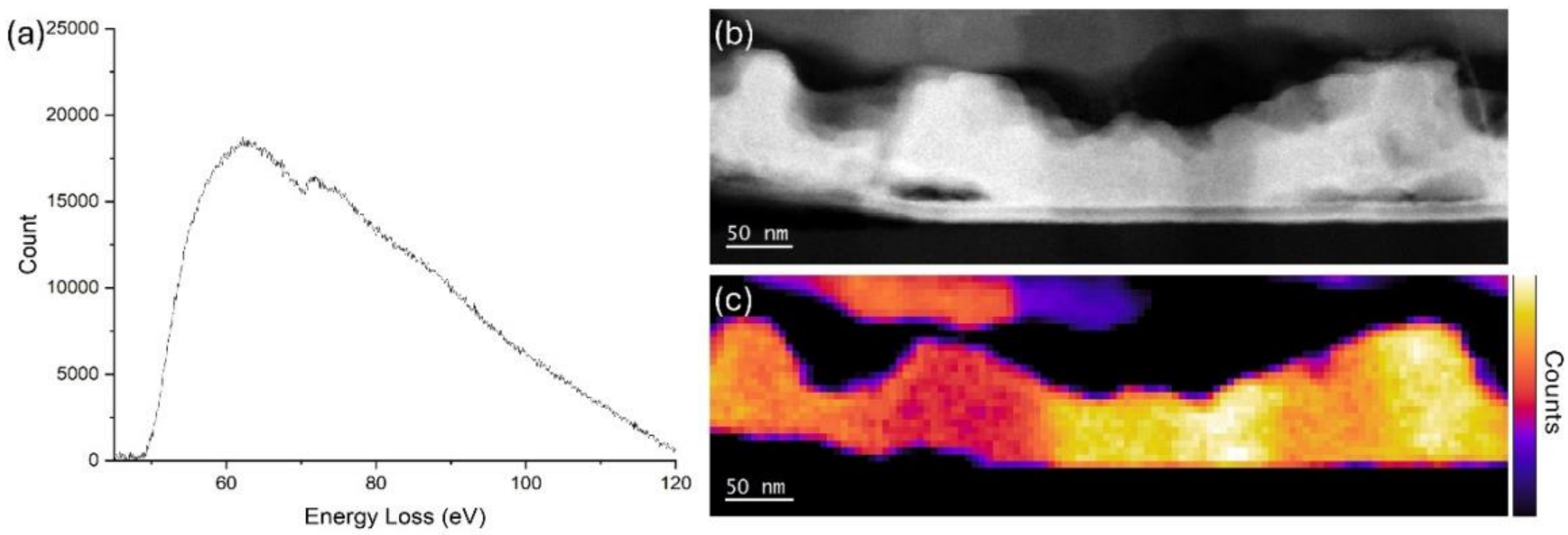

**Fig. 11:** (a) Cryo-EELS spectrum collected from the electrode region. (b) Cryo-HAADF-STEM image of the ROI, with corresponding EELS signal mapped onto the ROI in (c).

## Nanoscale Compositional Analysis of the Electrode via cryo-APT

Figure 12 (a) shows the full 3D reconstruction of the cryo-APT specimen prepared in Figure 8 (d). The reconstructed volume can be divided into three distinct regions: a Cr protective coating at the top, a central electrode region containing Pt and Li, and a lower region dominated by Si. Focusing on the electrode, Figure 12 (b) presents 2D contour plots of the relative concentrations of Li- and Pt-containing species in the X–Z orientation. Within the analysed volume, the detected ion counts of Li (14,412) and Pt (14,984) are low but nearly equal, consistent with the expected stoichiometry of the LiPt phase. Li- and Pt-containing species appear in close proximity but do not strongly overlap spatially, instead forming adjacent regions. Figure 12 (c) provides a 1D concentration profile along the Z-axis of the electrode, revealing two compositional regimes.

In Region 1, a clear separation between Pt-rich and Li-rich zones is observed at the interface, indicating significant compositional inhomogeneity and potential nanoscale segregation. This heterogeneity is evident from the steep decline of Pt concentration from nearly 100-atom % to below 20-atom % along the 1D profile, accompanied by a sharp increase in Li concentration. Such compositional gradients at electrochemical reaction fronts are consistent with phenomena reported in alloying anodes where lithium diffusivity limitations and mechanical stresses generate dynamic, compositionally diffuse interfaces[66, 67]. These local heterogeneities likely emerge from kinetic competition between lithium incorporation flux and platinum atom rearrangement near the SEI, facilitating strain accommodation and volume expansion mitigation during lithiation[68, 69].

By contrast looking at Region 2, located deeper within the electrode, the 3D reconstruction shows a more homogeneous spatial distribution of Pt and Li ions. The compositional maps and profiles indicate reduced nanoscale segregation or clustering, consistent with a relatively uniform alloying state in this region, with a composition close to 50:50 that corresponds to the LiPt phase identified in the XRD data in Figure 7 (a). Based on the established phase diagram of the Li–Pt system, the stoichiometric LiPt phase is expected to be an ordered intermetallic line compound under equilibrium conditions exhibiting long-range atomic order between Li and Pt ions[70, 71]. While difficult to directly resolve atomic ordering with the APT data, the observed compositional uniformity in Region 2 supports the presence of a well-mixed alloy phase without significant nanoscale compositional segregation. The corresponding 3D reconstructions for these two regions are shown in Figure 12 (d).

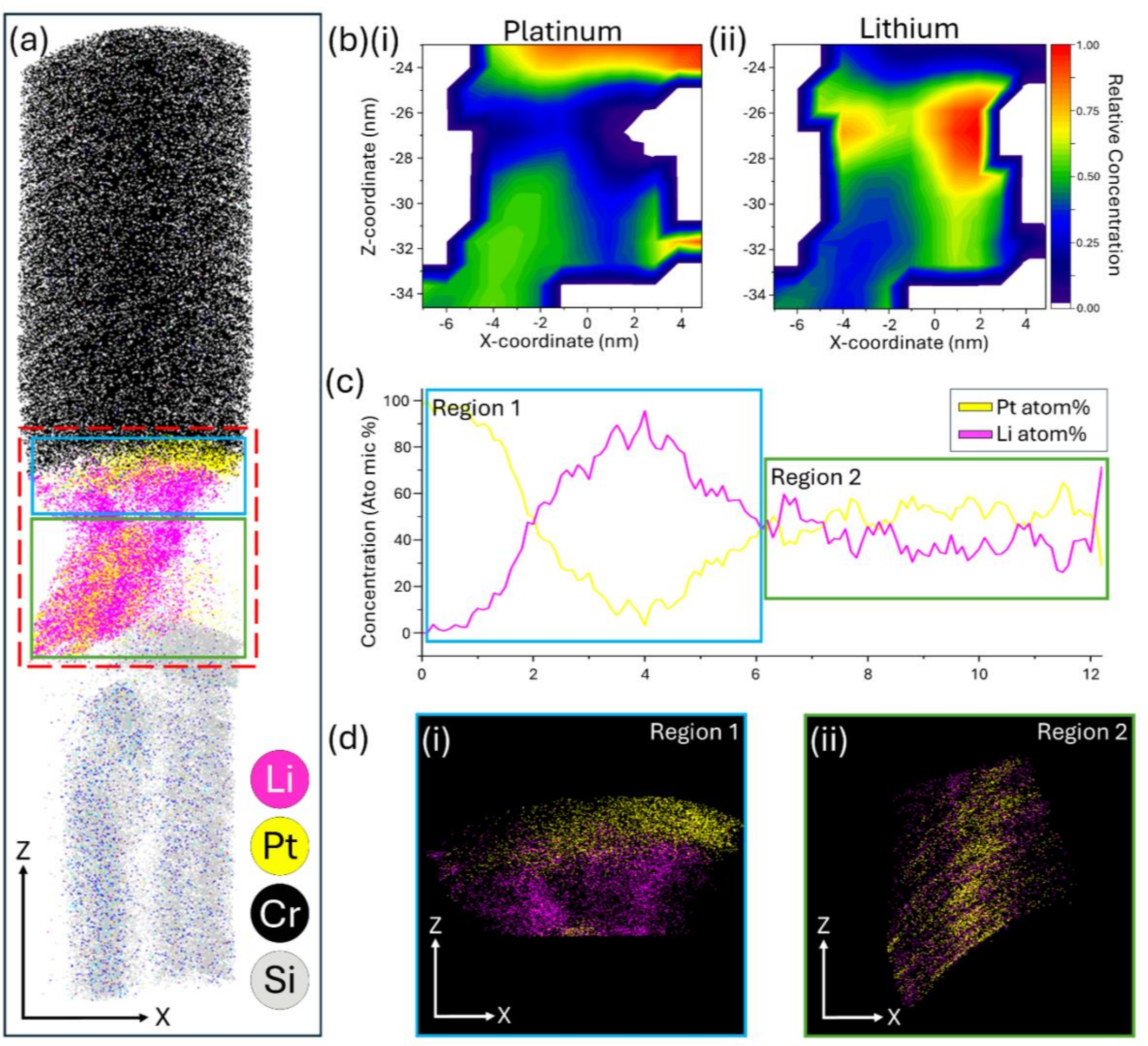

**Fig. 12:** (a) Full 3D cryo-APT reconstruction, highlighting the Cr protective coating (black), Li (pink) and Pt (yellow) from the LiPt electrode and Si (grey). (b) 2D contour plot showing the relative concentration of (i) Li-containing and (ii) Pt-containing species with the red serrated square in (a). (c) 1D concentration profile along the Z-axis of the electrode, showing two compositional regions, highlighted in blue (Region 1) and green (Region 2) in (a). (d) Isolated reconstructions of both regions showing (i) the spatial separation between Pt-rich and Li-rich zones in region 1 and (ii) the more homogeneous Li–Pt distribution deeper within the electrode.

Compositional gradients of Li and host metal species are a well-established feature of alloying anodes and reflect their intrinsic reaction kinetics[72]. Similar Li-metal gradients have been reported in Si, Sn, and Al systems, where non-equilibrium lithiation produces Li-rich and metal-rich domains governed by stress–composition coupling and limited Li diffusivity[72]. These gradients define dynamic reaction fronts that progressively homogenise with cycling as diffusion equilibrates, enabling strain relaxation and mitigating fracture. The present observation of a Pt-Li gradient near the interface in Region 1 and a compositional uniformity in the underlying bulk in Region 2 is therefore consistent with these established alloying behaviors, representing kinetically distinct regimes within the same electrode: a Li-flux-limited interfacial zone and a diffusion-controlled, well-mixed bulk. The coexistence of these kinetically distinct regimes facilitates the solid-solution type reaction mechanism observed during cycling. The Li-flux-limited interfacial zone, with its steep compositional gradients, represents the dynamic reaction front where the local rate of Li incorporation at the interface outpaces its diffusion into the alloy, leading to transient Li enrichment and compositional inhomogeneity. Meanwhile, the diffusion-controlled, well-mixed bulk provides a structurally stable environment where Li and Pt are homogeneously mixed, supporting reversible alloying and electrochemical stability. This partitioning between interface and bulk phases enables efficient strain accommodation and prevents mechanical failure, critical for the sustained performance of alloying anodes when taken together with the stabilised LiF-rich SEI. The observation of this behaviour at high resolution length scales has rarely been directly visualised, owing to Li's high mobility and beam-sensitive nature, underscoring cryo-APTs importance as a high resolution characterisation technique for next-generation battery anodes.

## Conclusion

This work provides a comprehensive framework that provides a multi-length scale view of the dynamic processes governing the performance of a Pt-based alloying anode. By combining in-situ synchrotron liquid cell electrochemistry XRF/XRD with cryo-STEM-EELS and cryo-APT, we directly tracked the sequence of lithiation–delithiation reactions, identifing the structural and chemical transitions underpinning long-term electrochemical stability. This combined approach links cycling-induced phase dynamics to near-atomic-scale interfacial chemistry and degradation mechanisms

In-situ XRD measurements reveal a discrete phase transformation during lithiation between the Pt anode and Li electrolyte to form $Li_2Pt$, followed by a mechanistic shift from discrete stepwise alloying to a solid-solution type reaction during delithation, resulting in a stable LiPt phase. These phase transformations were accompanied by the dynamic evolution of the SEI from an unstable, carbonate-rich state to a stable, LiF-dominated composition with cycling. Despite mechanical stresses and electrode redistribution during cycling revealed by XRF mapping, the persistence of the LiPt phase and LiF-rich SEI layer ensured sustained electrochemical reversibility.

Correlative cryogenic microscopy provided nanoscale insight into these processes. Cryo-STEM and EELS mapping confirmed the presence of a dense LiF-rich SEI, while cryo-APT revealed distinct compositional regimes within the electrode: a Pt–Li compositional gradient at the interface, reflecting Li-flux-limited kinetics, and a uniform LiPt alloy deeper in the bulk, consistent with diffusion-controlled equilibrium behaviour. These results provide direct experimental evidence for reaction-front–driven compositional heterogeneity and its gradual homogenisation during cycling.

Bringing together insights from the correlative measurements, this study highlights the critical interplay between interphase evolution, alloy phase stability, and electrochemical performance. The complexity and coupling of these nanoscale processes underscore the necessity of correlative, multi-technique methodologies. Beyond this Pt model system, the framework presented here offers a broadly applicable strategy for bridging in-situ structural dynamics with atomistic interfacial chemistry, providing essential mechanistic insight for the rational design of durable alloy-based electrodes in next-generation energy storage systems.

## Acknowledgments

N.M., M.P.R., M.S.C. acknowledge funding from Engineering and Physical Sciences Research Council (EPSRC) and Shell for funding through the InFUSE Prosperity Partnership (EP/V038044/1).

This work was made possible by the EPSRC Cryo-Enabled Multi-microscopy for Nanoscale Analysis in the Engineering and Physical Sciences EP/V007661/1. M.S.C. M.P., and Y. L. acknowledges funding from Royal Society Tata University Research Fellowship (URF\R1\201318) and Royal Society Enhancement Award RF\ERE\210200EM1. M.S.C. acknowledges funding from ERC CoG DISCO grant 101171966. SRJ thanks PhD funding from the Faraday Institution, under the grant EP/S514901/1. T.S. acknowledges support from EPSRC NAME Programme Grant EP/V001914/1. M.P.R acknowledges support from the Armourers and Brasiers Company. We wish to acknowledge the use of the EPSRC funded Physical Sciences Data-science Service hosted by the University of Southampton and STFC under grant number EP/S020357/1. The authors acknowledge beamtime and data analysis support from Diamond Light Source on beamline I14 under proposal number MG39828. The authors used OpenAI's ChatGPT to assist with code development in this work. All outputs were verified and validated by the authors.

# Supplementary Information

# Quantifying Phase Transformations in Alloying Anodes via In-Situ Liquid Cell Hard X-ray Spectroscopy and Cryogenic Microscopy

Neil Mulcahy[a], Syeda Ramin Jannat[a], Yaqi Li[a], Tigran Simonian[a], Mariana Palos[a], James O. Douglas[a], Jessica Walker[b], Baptiste Gault[a,c], Mary P. Ryan[a], Michele Shelly Conroy[a]

a. Department of Materials and London Centre for Nanotechnology, Imperial College London, Exhibition Road, London SW7 2AZ, U.K.

b. Diamond Light Source, Harwell Science and Innovation Campus, Didcot, UK

c. Max-Planck Institute for Sustainable Materials, Max-Planck Str. 1, 40237 Düsseldorf, Germany

d. Univ Rouen Normandie, CNRS, INSA Rouen Normandie, Groupe de Physique des Matériaux, UMR 6634, F-76000 Rouen, France

*Corresponding author: mconroy@imperial.ac.uk

## Description of code for analysing XRD and XRF data

The following Jupyter notebooks are provided as supplementary files to accompany this publication. Each notebook contains the code used for processing and analysing the XRD and XRF datasets described in the manuscript. A brief description of the purpose and functionality of each notebook is given below.

#### C1_XRFIntensitymaps.ipynb

Processes and visualises X-ray fluorescence (XRF) data. This includes:

- Reads element-specific XRF intensity data (Pt Lα) and associated spatial coordinates from processed NeXus (.nxs) files using the h5py library.
- Normalises all datasets to the global maximum intensity to enable consistent spatial comparison across multiple samples.
- Generates composite figures displaying normalised elemental intensity maps arranged in a grid layout, with shared colour scales and an integrated colourbar.
- Exports high-resolution images of the compiled XRF maps

#### C2_XRDBackgroundSubtraction.ipynb

Background correction of raw X-ray diffraction (XRD) intensity data. This includes:

- Imports raw XRD data from CSV files containing scattering vector (Q) values and measured intensities.
- Clips the dataset to exclude low-Q data ($Q < 1$ Å$^{-1}$) to minimize noise effects.
- Applies an asymmetric least squares (ALS) smoothing algorithm for baseline estimation and subtraction, implemented via sparse matrix operations
- Generates diagnostic plots displaying the original signal, estimated baseline, and baseline-corrected intensity to verify preprocessing quality.
- Exports the baseline-corrected data as CSV files for subsequent peak fitting and phase identification.

**C3_XRD_Simulations.ipynb**

Processes crystal structure information and generates simulated X-ray diffraction (XRD) patterns with intensities. This includes:

- Parsing crystallographic information files (CIFs) using the pymatgen library to extract atomic structures.
- Utilising pymatgen's XRDCalculator to compute diffraction peak positions (2θ), d-spacings, corresponding reciprocal space vectors (q), and normalized intensities over a specified angular range.
- Compiling peak data, including Miller indices (h, k, l), d-spacing, 2θ, q-values, and calculated intensities, into a pandas DataFrame.
- Exporting the compiled diffraction peak data to an Excel spreadsheet.

Experimental CIF files were acquired from Physical Sciences Data-science Service (PSDS)(www.psds.ac.uk)

**C4_XRD_PeakMatching.ipynb**

Performs phase identification by matching experimentally observed peak positions with simulated diffraction data. This includes:

- Loading experimentally fitted peak centers and simulated diffraction peak datasets from Excel files.
- Implementing a peak-matching algorithm with a specified tolerance in reciprocal space ($q \pm 0.01$ Å$^{-1}$) to identify candidate phases corresponding to each experimental peak.
- Annotating the experimental dataset with matched material phases based on the simulation, enabling direct comparison and phase assignment.
- Outputting the annotated experimental peaks as a CSV file for further analysis.

## Additional Materials and Methods

A Melbuild vacuum transfer and cryo cooling holder was used at liquid nitrogen temperature. A Thermo-Fischer Scientific (Waltham, Massachusetts, United states) Spectra 300 (S)TEM at 300 kV accelerating voltage was used for all STEM measurements. This instrument is probe corrected and fitted with an ultra-high-resolution X-FEG Ulti-monochromator. Nanobeam diffraction datasets were acquired using a 1.25 mrad probe convergence angle over a 64 × 64 pixel region with a 25 nm step size. The diffraction patterns were recorded on a Gatan K3 direct electron detector and summed using Gatan DigitalMicrograph GMS3 software.

In situ LCTEM experiments were conducted using the Stream system (DENSsolutions B.V., Delft, The Netherlands), involving the liquid cell TEM holder and a pressure-controlled liquid supply system (LSS). The MEMS nanochips utilised were similar in design to those used in the beamline setup and featured a three Pt electrode bottom biasing nanochip. The same commercial lithium electrolyte, $LiPF_6$ in ethylene carbonate/dimethyl carbonate (EC/DMC), (Merck Life Science UK Ltd, Dorset, UK), was circulated through the assembled TEM cell as in the X-ray experiments. Flow was regulated via the LSS and the Impulse software (DENSsolutions B.V., Delft, The Netherlands), with inlet and outlet pressures set to 2000 mbar and −950 mbar, respectively, resulting in a flow rate of approximately 8 μL min$^{-1}$.

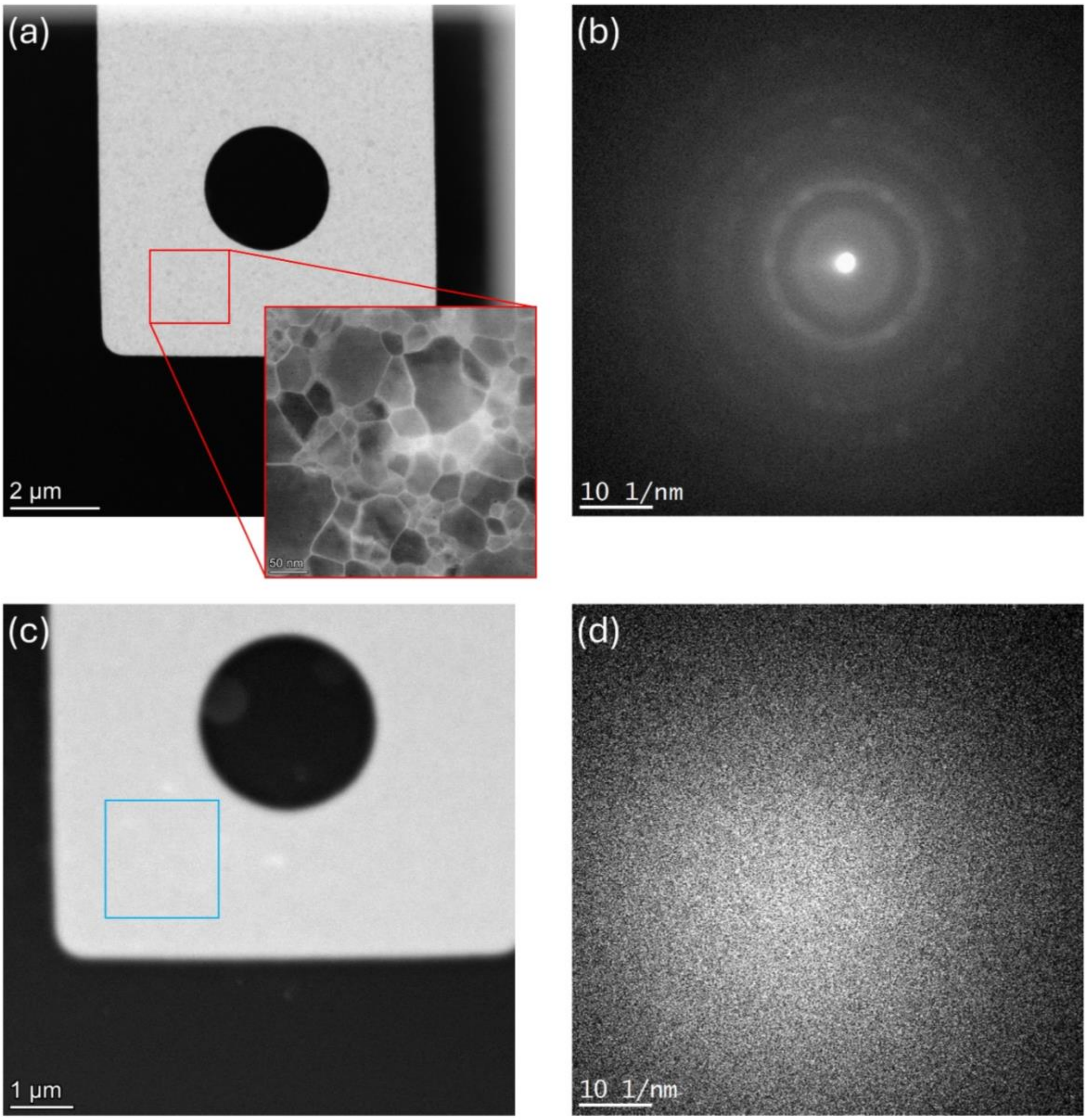

**Fig. S1:** High angle annular dark field scanning transmission electron microscopy (HAADF-STEM) images and electron diffraction patterns of a nanocell with and without electrolyte. (a) A HAADF-STEM image of the dry electrode. The inset image shows the polycrystalline nature of the Pt electrode, showing grains of varying sizes and orientations. (b) corresponding electron diffraction pattern captured from the red box in (a), with distinct diffraction rings visible confirming the polycrystalline structure. (c) HAADF-STEM image of the electrode after the addition of electrolyte, showing reduced image resolution and poorly resolved grains. (d) diffraction pattern from the blue box in (c), showing no clear features, illustrating the challenges of high-resolution imaging and diffraction in liquid environments using electron microscopy.

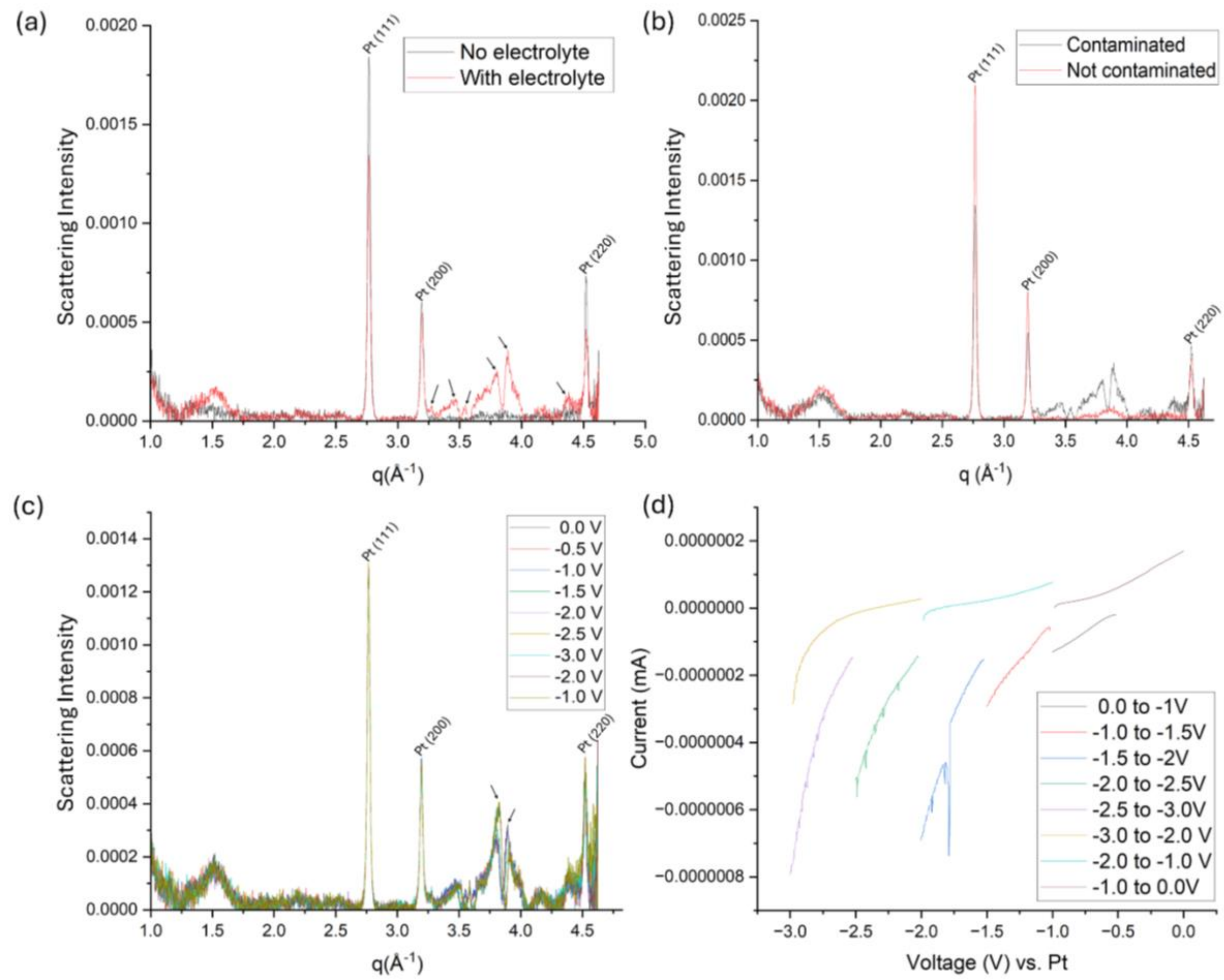

**Fig. S2:** (a) XRD dataset of an example of an assembled nanocell showing contamination. The black line corresponds to the cell without electrolyte, where distinct Pt reflections, Pt (111), Pt (200), and Pt (220), are clearly identified. The addition of electrolyte introduces sharp peaks between q = 3.25–4 $Å^{-1}$, indicating unexpected crystallinity likely due to contamination.
(b) Comparison of two XRD datasets: the red line represents an uncontaminated nanocell filled with electrolyte, while the black line shows a contaminated sample. The contaminated sample exhibits sharp additional peaks not present in the uncontaminated sample. (c) Multiple XRD datasets of the contaminated sample taken at various applied voltages. The crystalline peaks between q = 3.25–4 $Å^{-1}$ shift during biasing, highlighted by the black arrows, indicating that the contamination responds dynamically to the electrochemical environment. This peak shifting effectively obscures any features and renders the cell unusable for the intended experiments.
(d) C.V. curve showing the applied voltage steps. Current holds were performed at each voltage step and an XRD dataset was acquired.

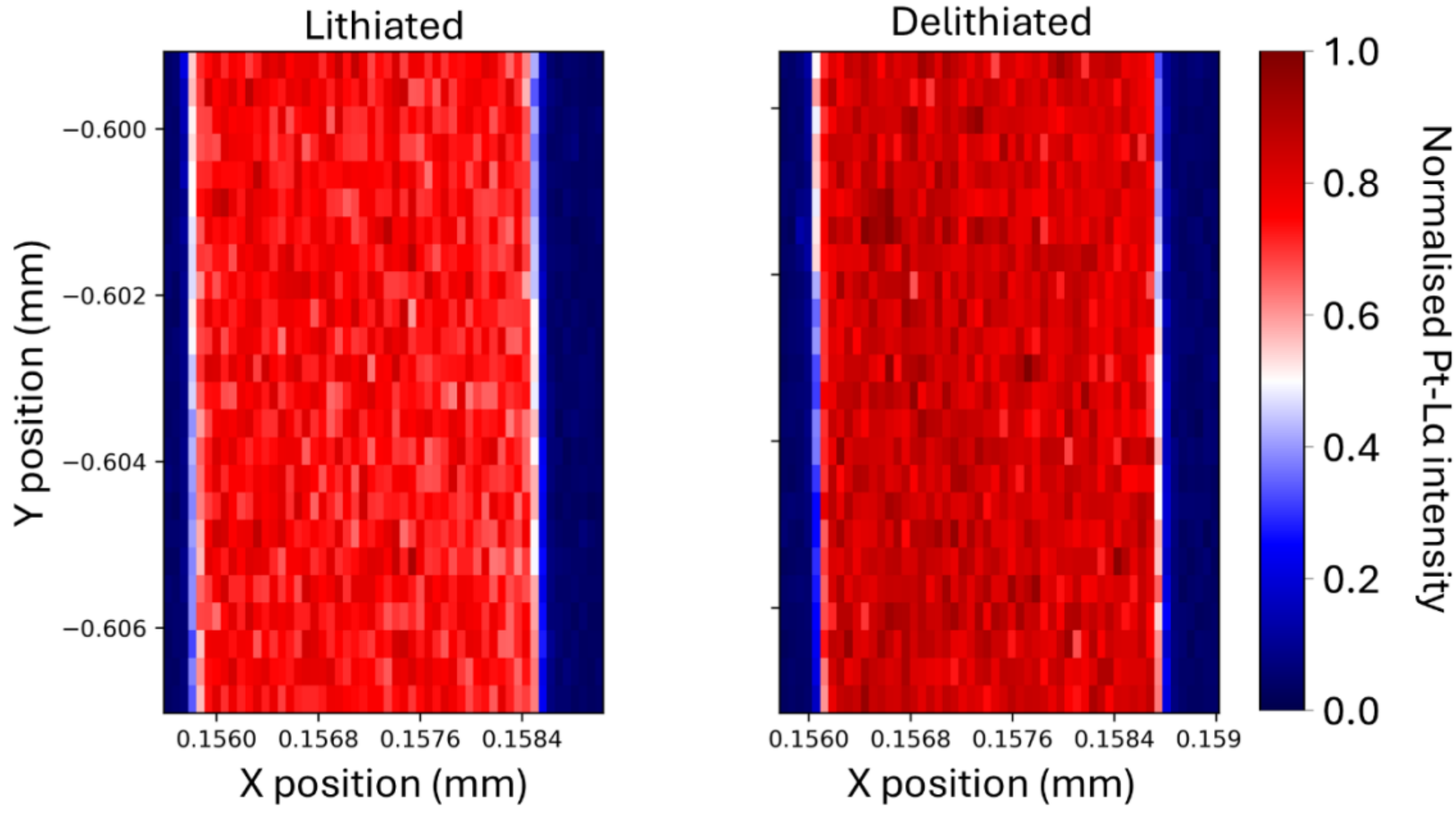

**Fig. S3:** XRF maps corresponding to the XRD patterns shown in Figure 5 (a) for the lithiated and delithiated electrode. A clear difference in overall intensity is observed between the two states, while the spatial distribution of Pt–Lα remains consistent.

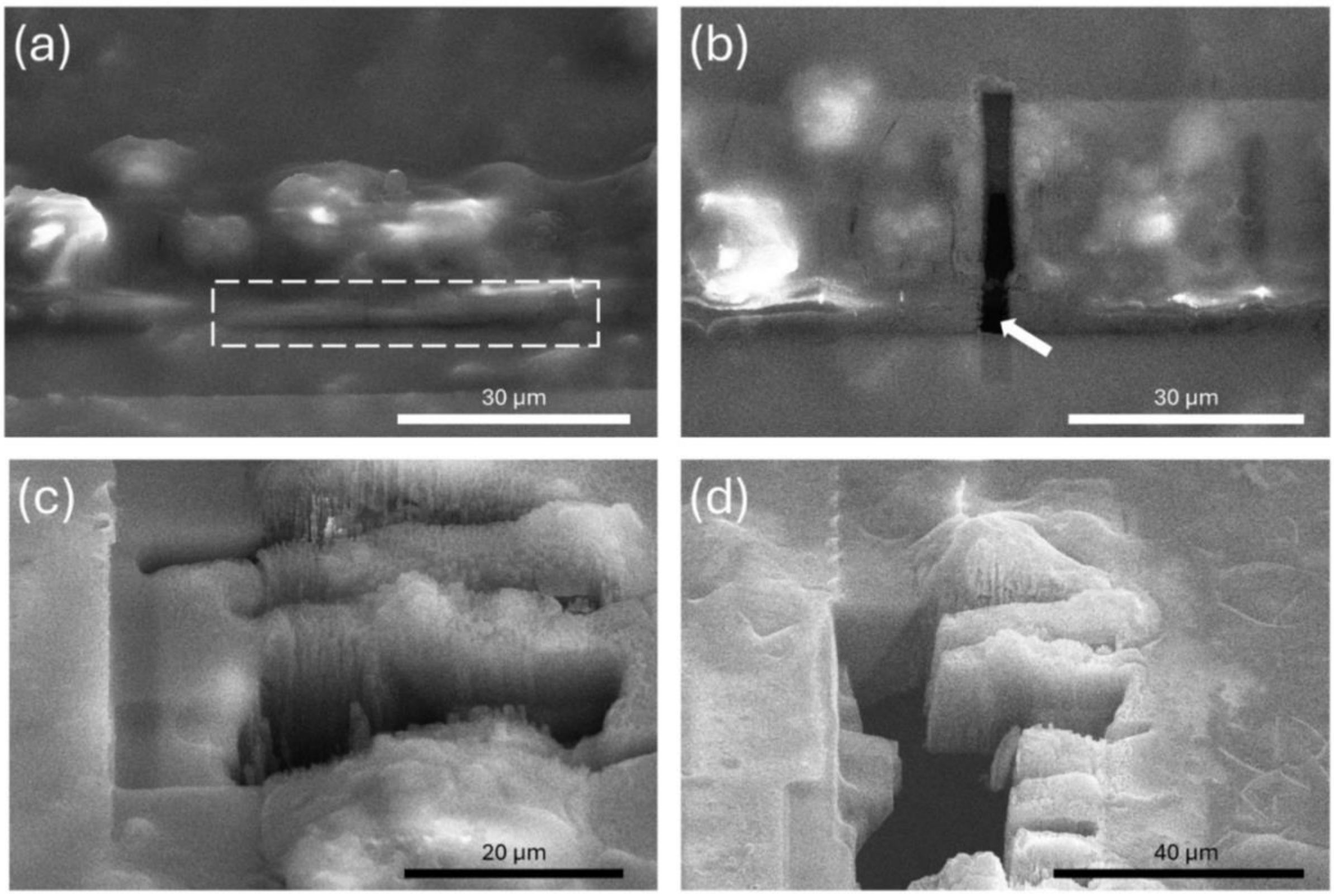

**Fig. S4:** SEM micrographs investigating the electrolyte covered area of the frozen MEMS nanochip. (a) Overview of the area covered by electrolyte at 52° stage tilt, showing a thick electrolyte layer and bulging of the $SiN_x$ membrane (highlighted by the serrated white square) (30 kV, 1.6 nA). (b) A hole milled through the SiNx window using Xe plasma (30 kV, 4 nA). The integrity of the window appears to be preserved, as the bottom of the viewing window is visible, highlighted by the white arrow (SEM, 30 kV, 1.6 nA). (c, d) Attempts to prepare lift-out samples from the electrolyte-covered region. The large volume of electrolyte made it challenging to precisely locate the electrodes and resolve features of interest on their surfaces. (SEM, 30 kV, 0.4 nA).

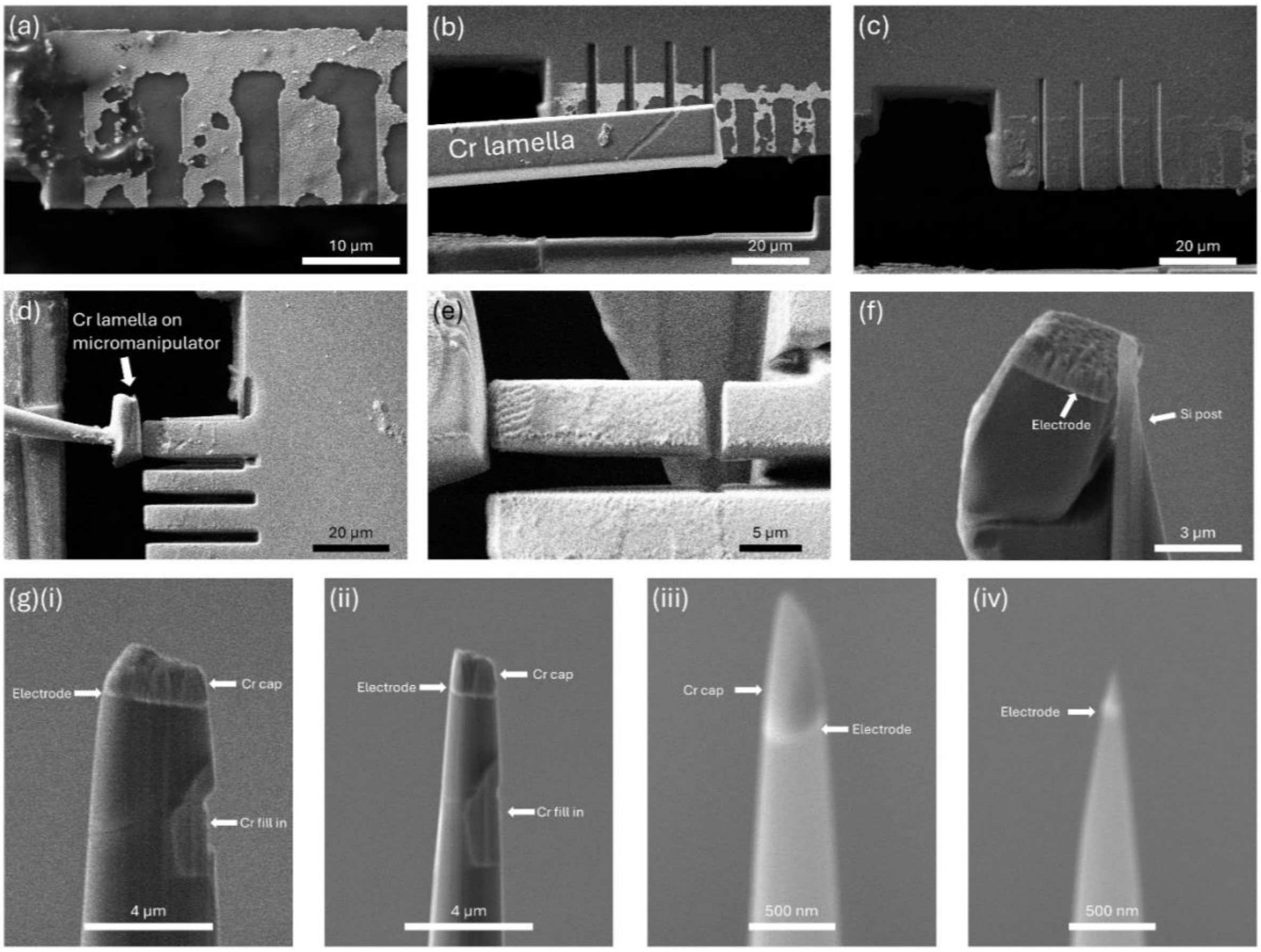

**Fig. S5:** SEM micrographs (10 kV, 0.4 nA) illustrating the full cryo-APT sample preparation workflow. (a) shows a number of partially uncovered electrodes on the MEMS nanochip. (b) A pre-prepared Cr lamella is brought over the ROI and the Xe ion beam (30 kV, 1 nA) is rastered across the lamella to deposit a thin site-specific Cr protection layer. (c) the resulting electrode region after Cr deposition. (d) Electrodes milled into lift-out bars using the Xe ion beam (30 kV, 1–4 nA). A Cr lamella and redeposition welding were used to attach one lift-out bar to the micromanipulator at cryogenic temperature. (e) Example of a sample successfully lifted out and attached to the micromanipulator. (f) an example of a lift out bar attached to a Si microarray post. Following this the underside of the sample is filled in with Cr through redeposition welding and the sample is milled to fit the post as in (g). In (g)(i) this fill in is evident along with the electrode and the Cr protection layer from (b) and (c). (g)(i-iv) tracks the final thinning process where the final analysed needle (Xe plasma, 30 kV, 30 pA- 1 nA) with preserved electrode is evident in (iv).

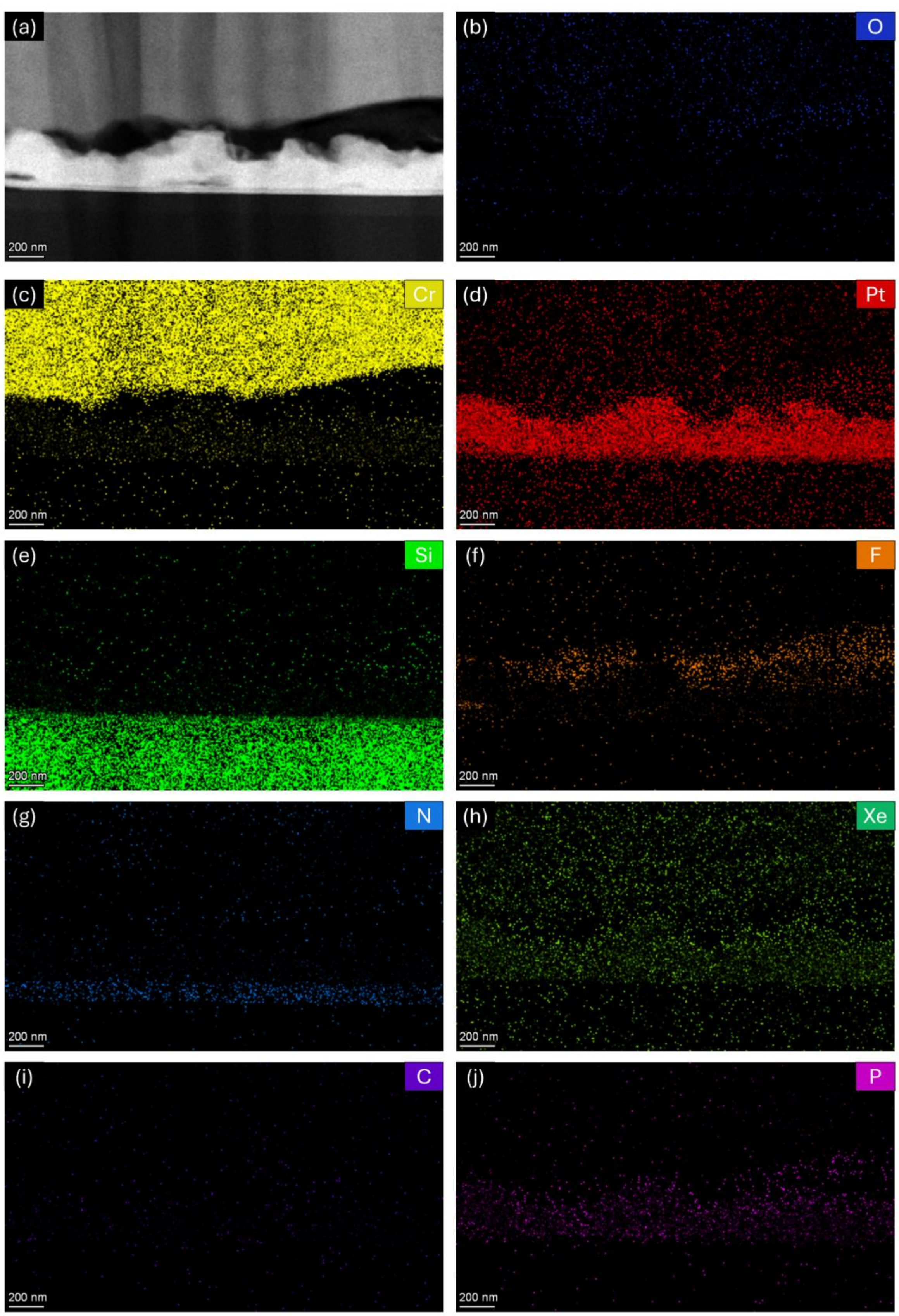

**Fig. S6:** (a) HAADF STEM image of ROI used for EDX analysis. (b-i) Individual EDX maps for O, Cr, Pt, Si, F, N, Xe, C and P. All scale bars are 200 nm.

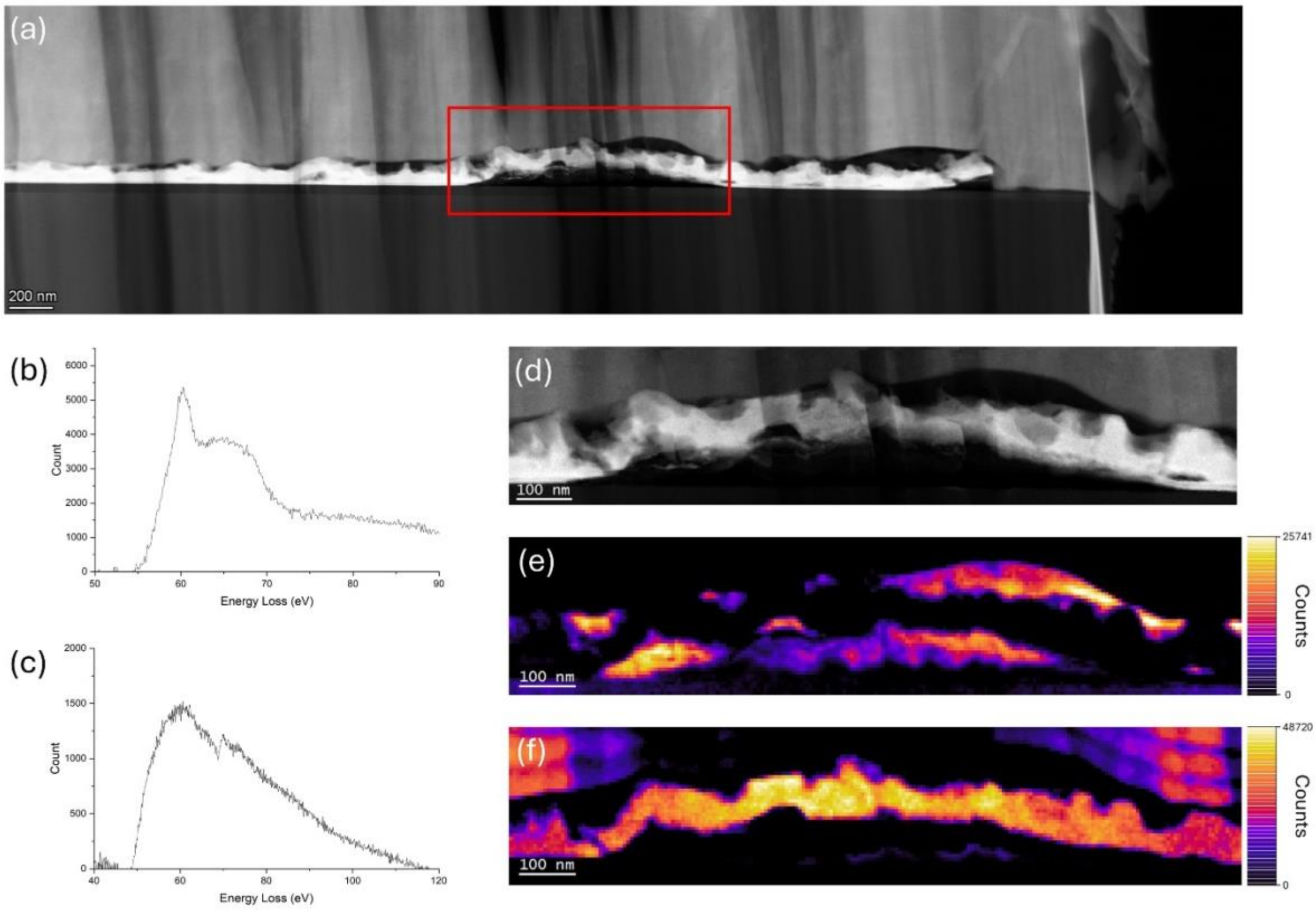

**Fig. S7:** (a) HAADF STEM image of an overview of the prepared lamella with an ROI with apparent delamination highlighted by red square, shown in (d). (b) EELS spectrum capturing SEI peaks, mapped in (e), and (c) EELS spectrum capturing Pt- O and Pt-N edges mapped in (f). A distinct SEI is visible above and below the electrode in this region, suggesting that the bottom surface detached from the SiNx membrane and was exposed to the electrolyte. Despite this delamination, the electrode remained electrochemically active, as evidenced by SEI formation on both surfaces.

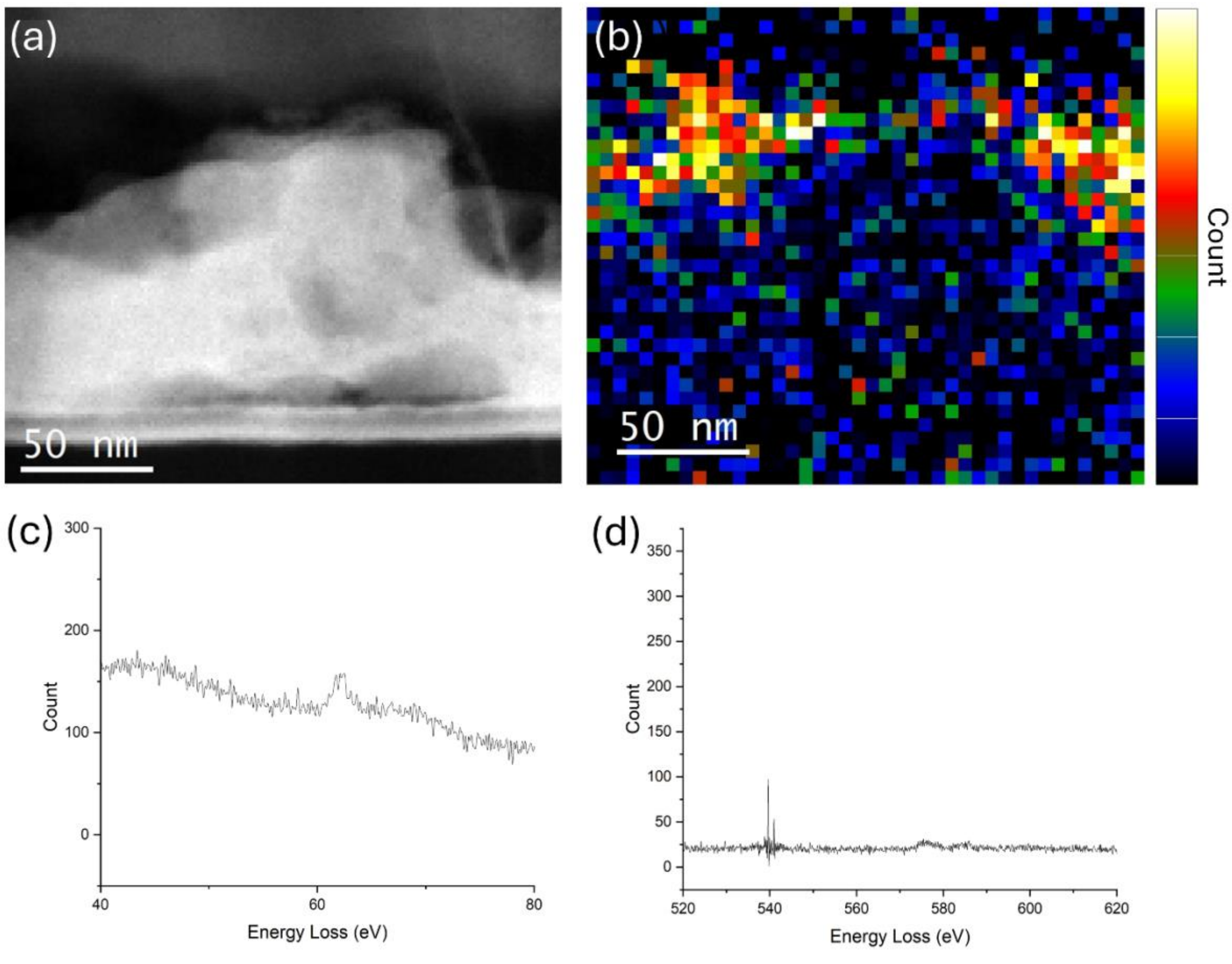

**Fig. S8:** Quantifying Oxygen content in SEI via EELS. (a) HAADF STEM image section of ROI presented in Figure 10. (b) EELS map corresponding to SEI EELS edge present in (c). (d) High loss EELS range capturing no Oxygen K-edges.

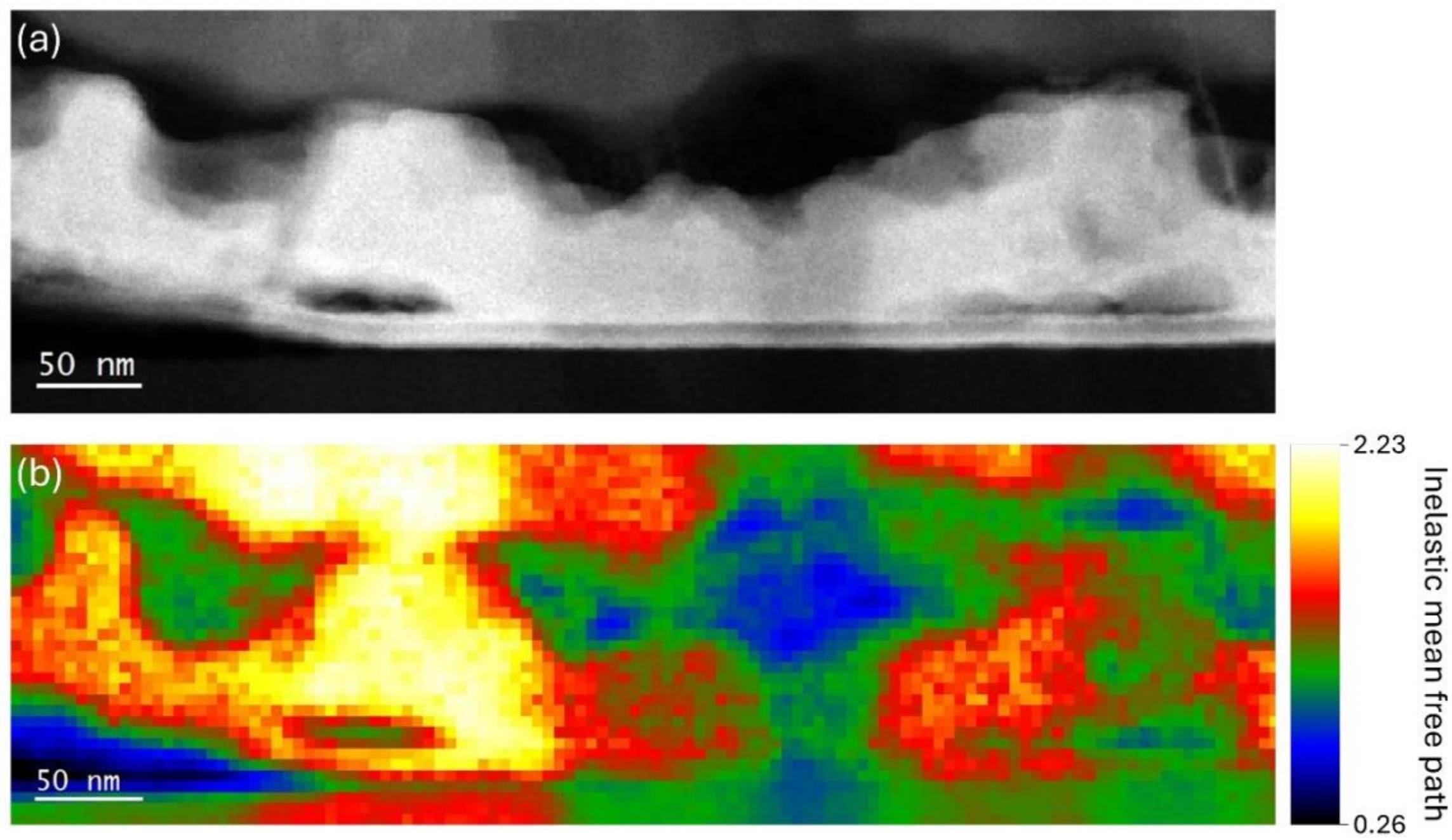

**Fig. S9:** (a) HAADF-STEM image of the select ROI containing cycled Pt alloy electrode and distinct SEI layer (see main text). (b) Relative thickness map of the same region, showing relative intensity of inelastic mean free path length (λ), demonstrating significant spatial variation in sample thickness across the SEI and electrode. Colour scale indicates λ values from 0.26 to 2.23, with thinner regions appearing blue and thicker regions in yellow/red.